\documentclass[a4paper,11pt]{article}
\pdfoutput=1 
\usepackage{jcappub} 
\usepackage[T1]{fontenc}

\title{\boldmath Weak lensing probe of cubic Galileon model}

\author[a,1]{Bikash R. Dinda,\note{Corresponding author.}}

\affiliation[a]{Centre for Theoretical Physics, Jamia Millia Islamia,\\New Delhi-110025, India}

\emailAdd{bikash@ctp-jamia.res.in}

\abstract{The cubic Galileon model containing the lowest non-trivial order action of the full Galileon action can produce the stable late-time cosmic acceleration. This model can have a significant role in the growth of structures. The signatures of the cubic Galileon model in the structure formation can be probed by the weak lensing statistics. Weak lensing convergence statistics is one of the strongest probes to the structure formation and hence it can probe the dark energy or modified theories of gravity models. In this work, we investigate the detectability of the cubic Galileon model from the $ \Lambda $CDM model or from the canonical quintessence model through the convergence power spectrum and bi-spectrum.}

\begin{document}
\maketitle
\flushbottom

\section{Introduction}

In literature, it is a well-known fact that the Universe is accelerating. This accelerating phase of the Universe is called late-time cosmic acceleration. Some observations like Supernova Type-Ia observations \citep{SnIa1,SnIa2}, cosmic microwave background observations \citep{CMB1,CMB2,Planck}, baryon acoustic oscillations measurements \citep{BAO1,BAO2} confirm this fact. This acceleration can be explained by introducing an exotic matter called dark energy, which has negative pressure. The best possible simple dark energy model is the $ \Lambda $CDM model. In this model at late time the Universe consists of cosmological constant, $ \Lambda $ (appx. 70$ \% $ at present) and cold dark matter (CDM) component (appx. 30$ \% $ at present) \citep{Planck}. However, the  $ \Lambda $CDM model has some serious theoretical problem like the fine-tuning problem (If we consider energy density of the cosmological constant ($\Lambda$) as the vacuum energy density, then the theoretical value associated with the vacuum energy density is nearly $10^{123}$ times the observed value of the energy density of the cosmological constant from different cosmological observations. this discrepancy is referred to the fine-tuning problem in the literature of cosmology.) and cosmic coincidence problem (The energy density of the cosmological constant is constant with the evolution of the Universe, whereas the energy density of the matter decreases with time. At present, the ratio of the energy densities of the cosmological constant and the matter is nearly $7:3$ (from different cosmological observations) and the cosmological constant dominates over matter recently at $z\approx0.3-0.5$ (depending on the cosmological model parameters). The question is why the dominance of the cosmological constant over matter is at a particular redshift or why at present their ratio is comparable. This extreme coincidence is sometimes referred as the cosmic coincidence problem.) \citep{fntn}. Despite these theoretical problems, although it is the best possible fit to the most of the cosmological observations, recently some observations \citep{inctny1,inctny2,inctny3,inctny4} suggest that it also has some inconsistencies to fit some data. This fact motivates us to go beyond the $ \Lambda $CDM model. Another alternative to explain the cosmic acceleration is the dark energy with evolving equation of state such as quintessence \citep{Tracker1,Tracker2,ThQ1,ThQ2,ThQ3,ThQ4,ThQ5,ThQ6} etc. Although very few evolving dark energy models like tracker models \citep{Tracker1,Tracker2} can be free from the cosmic coincidence problem, the fine-tuning problem is not avoidable.

Another popular alternative to the dark energy models is the modified theory of gravity \citep{mog1,mog2,mog3,mog4,mog5}, which can explain the late time cosmic acceleration. One of the modified theories of gravity is the Galileon gravity \citep{galBasic1,galBasic2,galBasic3,galBasic4,galBasic5,galBasic6,galBasic7,galBasic8,galBasic9,galBasic10,galfull1,galfull2,galfull3,galfull4,galfull5,galfull6,galfull7}. The Galileon theory is obtained by taking the decoupling limit of the Dvali-Gabadadze-Porrati (DGP) \citep{DGP}. In Galileon theory modification to the general relativity arises through a Galileon invariant scalar field, called Galileon. It is invariant under the Galileon shift symmetry $ \partial_{\mu} \phi \rightarrow \partial_{\mu} \phi + b_{\mu} $ in 4-D Minkowski spacetime where $ b_{\mu} $ is a constant vector. This Galileon shift symmetry confirms that the equation of motion remains second order in field derivatives despite the presence of the higher derivative terms in the action \citep{galBasic1,galfull2,galfull7} and the Galileon theory is free from Ostrogradsky ghosts \citep{OstrgGhst}. The Galileon theory is a subset of the general Horndeskii theory \citep{Hrnski}. Although the shift symmetry breaks down in general cosmological metric like Friedmann-Robertson-Walker (FRW) metric, the equation of motion still remains second order by adding suitable couplings between Galileon derivatives and curvature tensors. Introduction of these types of coupling terms makes the modification to the general relativity highly non-trivial.

Due to the presence of such non-trivial terms, the modified theories of gravity possess the extra degree(s) of freedom which can affect the local physics which is extremely consistent with the general theory of relativity \citep{GRsolarsys}. This extra degree of freedom is responsible for the 'fifth force'. This fifth force should be highly suppressed on the local scale such as in solar system as because on the local scale any modification to the Einstein's general theory of relativity is highly constrained \citep{GRsolarsys}. The Vainshtein mechanism \citep{Vainshtein} is one of the mass screening methods which preserve the local physics in the modified theories of gravity models by suppressing the fifth force on local scales. In 1972 Vainshtein invented this method to address the van Dam-Veltman-Zakharov (vDVZ) discontinuity \citep{vDVZ1,vDVZ2} in the linear theory of massive gravity of Pauli-Fierz \citep{MsGPF}. In general the Galileon gravity action contains five terms denoted by $ L_{1} $ to $ L_{5} $. The Vainshtein mechanism is based on the term like $ L_{3} = (\partial_{\mu} \phi)^{2} \Box \phi $ \citep{gal7,galfull8}. This is the lowest non-trivial term required for the mass screening in Vainshtein mechanism. $ L_{4} $ term can take part in the mass screening but does not introduce any new physics \citep{gal7}. $ L_{5} $ term does not contribute to the mass screening \citep{gal7}. In this paper, we consider up to the third terms. In literature generally, this is called cubic Galileon model. The second term is the usual canonical kinetic term. The second and third terms together only cannot give stable late time acceleration \citep{galfull3}. In addition to the second and third terms, the first term which is linear in Galileon field can produce stable acceleration. So, in Galileon theory the cubic Galileon model has the lowest non-trivial action which can produce stable late-time cosmic acceleration \citep{gal1,gal2,gal3,gal4,gal5,gal6,gal7}.

One of the major goals of the recent and upcoming high precision cosmological observations is to determine whether the late time cosmic acceleration is due to the $ \Lambda $CDM model or any evolving dark energy model or any modified theory of gravity by studying the signatures of these models on both the background evolution of the Universe and the growth of the structures. Among these weak lensing experiments \citep{WLp1,WLp2,WLp3,WLp4,WLp5,WLp6,WLp7,WLp8,WLp9,WLp10,WLp11,WLp12,WLp13,WLp14} are particularly promising to determine the nature of dark energy. In this paper, we address the prospect of probing cubic Galileon model in light of the weak lensing measurements by convergence statistics. Weak lensing is the statistical measure of the image distortion effect of the distant background galaxies due to gravitational bending of light by the intervening large scale structures along the photon geodesic. The weak lensing around massive halos was first measured in the nineties \citep{tyson,brainerd} but the first measurement of weak lensing due to large scale structures was done independently by four groups in 2000 \citep{WLSF1,WLSF2,WLSF3,WLSF4}. After that weak lensing become one of the strongest probes to the large scale structure formation. The main advantage of the weak lensing is that it solely depends on the underlying total matter-energy overdensity and hence complicated bias modeling can be avoided. Present and upcoming surveys like DES \citep{des}, LSST \citep{lsst}, Euclid \citep{euclid}, WFIRST etc. can provide us accurate weak lensing measurements which in turn can give us an accurate idea whether the late time cosmic acceleration is due to the $ \Lambda $CDM model or any evolving dark energy model or any modified theory of gravity model.

Previously, in the literature, the effect of modified gravity has been studied in some particular models \citep{pwlnew3,wlMOG1,wlMOG3,wlMOG4,wlMOG5,wlMOG6,wlMOG7,wlMOG8}. In \citep{pwlnew3,wlMOG1}, weak lensing effect has been studied mainly in the $f(R)$ gravity models. In \citep{wlMOG3}, signatures of the modified gravity models have been studied in general by introducing degeneracies in the modification to the gravity. In \citep{wlMOG4} modified gravity models have been studied in the model independent way in the context of weak lensing in the COSMOS survey. In \citep{wlMOG5,wlMOG6,wlMOG7}, $f(R)$ gravity and DGP \citep{DGP} type models have been investigated through weak lensing. In \citep{wlMOG8}, weak lensing has been studied in the context of scalar-tensor theories and $f(R)$ gravity as a special case. However, scalar-tensor type theories have been ruled out by the GW170817 result \citep{gwMOG1,gwMOG2,gwMOG3,gwMOG4} from the speed of the propagation of the gravitational waves. This is because of the fact that the speed of the gravitational waves deviates from $C$ when tensor modes are coupled with the other modes like in the non-minimally coupled scalar field models. Although the full Galileon gravity (due to $L_{4}$ and $L_{5}$ terms) is non-minimally coupled to gravity, the cubic Galileon gravity is minimally coupled. So, the cubic Galileon model is still interesting to study its signatures on the large-scale structure observables like in the weak lensing convergence power spectrum, bispectrum etc.

Keeping these facts in mind, in this paper we aim to detect the signatures of the cubic Galileon model from the $ \Lambda $CDM model or from the quintessence models using weak lensing statistics. We work in the unit, $ C= \hbar =1 $ and we consider ($-$,+,+,+) signature throughout the paper.

The paper is organised as: in section 2 we briefly discuss the action of the cubic Galileon model; in section 3 background evolution has been studied; in section 4 we discuss the evolution of the perturbation in the cubic Galileon model; in section 5 we present the signature of the cubic Galileon model on the convergence power spectrum and bi-spectrum; and finally, in section 6 we present our conclusion.

\section{Action of the model}

We consider cubic Galileon model to study the evolutionary history of the Universe. The action for the cubic Galileon field, $\phi$ is given by \citep{gal1,gal2,gal3,gal4,gal5,gal6,gal7}

\begin{equation}
S=\int d^4x\sqrt{-g}\Bigl [\frac{M^2_{\rm{pl}}}{2} R +  \frac{1}{2} \sum_{i=1}^{3} c_{i} \mathcal{L}_{i} \Bigr] + \mathcal{S}_m \, ,
\label{eq:actionmain}
\end{equation}

\noindent
where $ \mathcal{L}_{1} = M^{3} \phi $, $ \mathcal{L}_{2} = (\nabla \phi)^2 $ and $ \mathcal{L}_{3} = \frac{(\nabla \phi)^2}{M^{3}} \Box \phi $. $\mathcal{S}_m$ is the action for the matter counterpart, $M_{pl}$ is the reduced Planck mass, M is a mass dimensional constant, $c_{i}^{,}s$ are dimensionless constants and g is the determinant of the metric describing the Universe. Note that full Galileon model has other two terms containing higher order derivatives of the Galileon field \citep{galfull1,galfull2,galfull3,galfull4,galfull5,galfull6}. Those two terms involve nonminimal coupling. But cubic Galileon model considered here has no non-minimal coupling.

Note that for the matter counterpart we take cold dark matter (CDM) and baryons together as the total matter. For simplicity we take $ c_{2} = - 1 $ and this choice does not change the essence of the cubic galileon model. We define $\frac{c_{3}}{M^{3}} = - \frac{\alpha}{M^{3}} = - \beta$. Now we define the linear term in the action \eqref{eq:actionmain} in a way that it looks like a potential given by $V(\phi) = - \frac{1}{2} c_{1} M^{3} \phi$. So, the action \eqref{eq:actionmain} looks like \citep{gal4,gal5,gal6,gal7}

\begin{equation}
S=\int d^4x\sqrt{-g}\Bigl [\frac{M^2_{\rm{pl}}}{2} R -  \frac{1}{2}(\nabla \phi)^2\Bigl(1 + \beta \Box \phi\Bigr) - V(\phi) \Bigr] + \mathcal{S}_m \, ,
\label{eq:action}
\end{equation}

The purpose to write down the action \eqref{eq:actionmain} in the form given in equation \eqref{eq:action} is that for $\beta = 0$ the action reduces to the standard quintessence action with linear potential \citep{ThQ1,ThQ2,ThQ3,ThQ4,ThQ5}.

\section{Background evolution}

For the background evolution of the Universe, we consider flat FRW metric given by $ds^{2} = - dt^{2} + a^{2} (t) d\vec{r}.d\vec{r}$ where $t$ is the cosmic time, $\vec{r}$ is the comoving coordinate vector and $a$ is the cosmic scale factor. Varying the action \eqref{eq:action} with respect to the metric the background Einstein equations become \citep{gal1,gal2,gal3,gal4,gal5}

\begin{equation}
3M_{\rm pl}^2H^2 = \bar{\rho}_{\rm m}+\frac{\dot{\phi}^2}{2}\Bigl(1-6\beta H\dot{\phi}\Bigr)+V{(\phi)},
\label{eq:first_Friedmann}
\end{equation}

\begin{equation}
M_{\rm pl}^2(2\dot H + 3H^2) = -\frac{\dot{\phi}^2}{2}\Bigl(1+2\beta \ddot{\phi}\Bigr)+V(\phi),
\label{eq:second_Friedmann}
\end{equation}
 
\noindent
where overdot is the derivative with respect to the cosmic time $t$, $H$ is the Hubble parameter and $ \bar{\rho}_{\rm m} $ is the background matter energy density. The background Euler-Lagrangian equation for the Galileon field $ \phi $ is given by \citep{gal1,gal2,gal3,gal4,gal5}

\begin{equation}
\ddot{\phi} + 3H\dot{\phi}-3\beta \dot{\phi}\Bigl(3H^2\dot{\phi}+\dot{H}\dot{\phi}+2H\ddot{\phi}\Bigr)+ V_{\phi}=0,
\label{eq:E-L_eq}
\end{equation}

\noindent
where subscript $\phi$ is the derivative with respect to the field $\phi$. Note that for the simplicity of the notation we take same $\phi$ as the background field. 

To study the background evolution, in literature, it is a common practice to rewrite the above differential equations with an autonomous system of equations. To do so first we define some dimensionless background quantities given by \citep{gal4,gal5}

\begin{eqnarray}
x &=& \frac{ \dot{\phi} }{\sqrt{6} H M_{Pl}} = \frac{\Big{(} \dfrac{d \phi}{d N} \Big{)}}{\sqrt{6} M_{Pl}}, \hspace{1 cm}  y = \frac{\sqrt{V}}{\sqrt{3} H M_{Pl}}, \nonumber\\
\epsilon &=& -6 \beta H \dot{\phi} = -6 \beta H^{2} \Big{(} \dfrac{d \phi}{d N} \Big{)}, \hspace{0.5 cm}  \lambda = - M_{Pl} \frac{V_{\phi}}{V}, \nonumber\\
\Gamma &=& V \frac{V_{\phi \phi}}{V_{\phi}^{2}} = 0\hspace{0.2 cm} (\text{Here}),
\label{eq:dimless_var}
\end{eqnarray}

\noindent
where $ N = lna $ is the e-folding. Note that the quantity $x$ here is different from the coordinate $x$ used in eq. \eqref{eq:actionmain}. In the subsequent sections by $x$ we mean the quantity defined in Eq. \eqref{eq:dimless_var}. Using these quantities the autonomous system of equations become \citep{gal4,gal5}

\begin{eqnarray}
\dfrac{d x}{d N} &=& \frac{3 x^3 \left(2+5 \epsilon +\epsilon^2\right)-3 x \left(2-\epsilon +y^2 (2+3 \epsilon )\right)+2 \sqrt{6} y^2 \lambda -\sqrt{6} x^2 y^2 \epsilon  \lambda }{4+4 \epsilon +x^2 \epsilon^2}, \nonumber\\
\dfrac{d y}{d N} &=& -\frac{y \left(12 \left(-1+y^2\right) (1+\epsilon )-6 x^2 \left(2+4 \epsilon +\epsilon^2\right)+\sqrt{6} x^3 \epsilon^2 \lambda +2 \sqrt{6} x \left(2+\left(2+y^2\right) \epsilon \right) \lambda \right)}{8+8 \epsilon +2 x^2 \epsilon^2}, \nonumber\\
\dfrac{d \epsilon}{d N} &=& -\frac{\epsilon  \left(-3 x \left(-3+y^2\right) (2+\epsilon )+3 x^3 \left(2+3 \epsilon +\epsilon^2\right)-2 \sqrt{6} y^2 \lambda -\sqrt{6} x^2 y^2 \epsilon  \lambda \right)}{x \left(4+4 \epsilon +x^2 \epsilon^2\right)}, \nonumber\\
\dfrac{d \lambda}{d N} &=& \sqrt{6}x\lambda^2(1-\Gamma),
\label{eq:auto_sys}
\end{eqnarray}

\noindent
where we have changed the time derivative to the derivative with respect to N. By using proper initial conditions (which are discussed in the next subsection), we solve the above-mentioned autonomous system of coupled differential equations to find the evolution of the background quantities.

Now, having solutions of the autonomous system of equations in Eq. \eqref{eq:auto_sys} and from the above-mentioned dimensionless quantities in \eqref{eq:dimless_var}, we get some important background quantities given below: \citep{gal4,gal5}

\begin{eqnarray}
\omega_{\phi} &=& \frac{3 x^2 (\epsilon  (\epsilon +8)+4)-2 \sqrt{6} \lambda  x y^2 \epsilon -12 y^2 (\epsilon +1)}{3 \left(\epsilon  \left(x^2 \epsilon +4\right)+4\right) \left(x^2 (\epsilon +1)+y^2\right)}, \nonumber\\
\Omega_{\phi} &=& x^2 (\epsilon +1)+y^2, \nonumber\\
\Omega_{m} &=& 1 - \Omega_{\phi}, \nonumber\\
H^{2} &=& H_{0}^{2} \frac{\Omega_{m}^{(0)} (1+ z)^{3}}{\Omega_{m}},
\label{eq:imp_bkg_qnt}
\end{eqnarray}

\noindent
where $w_{\phi}$ is the equation of state of the Galileon field, $\Omega_{\phi}$ and $\Omega_{m}$ are the energy density parameters of the Galileon field and total matter respectively,  $H$ is the Hubble parameter, $H_{0}$ and $\Omega_{m}^{(0)}$ are the present day Hubble and matter-energy density parameters respectively and $z$ is the redshift.

\subsection{Initial conditions}

First of all we fix the initial conditions of all the four quantities ($ x $, $ y $, $ \epsilon $ and $ \lambda $ in Eq. \eqref{eq:auto_sys}) at redshift $ z = 1100 $ in early matter-dominated era. At this redshift, we can neglect any contribution from the dark energy. In literature, we generally consider two types of initial conditions for these types of cosmological models. One is the tracker condition \citep{Tracker1,Tracker2} and another is the thawing condition \citep{ThQ1,ThQ2,ThQ3,ThQ4,ThQ5,ThQ6}. In tracker models, the scalar field tracks the background initially and at late times the equation of state of the scalar field freezes to a constant value near to $ -1 $. In thawing class of models, the scalar field is initially frozen to a value $ w_{\phi} \approx -1 $ due to the large Hubble friction in the early matter-dominated era. At late times the scalar field thaws away from its initial frozen state and the equation of state of the scalar field goes to higher values ($ w_{\phi} > -1 $) accordingly. In references \citep{ThQ3,ThQ7,ThQ8}, we can see that for the polynomial potential of the form $V(\phi) \propto \phi^{n}$, to produce a tracker type initial condition $n<0$ is required. To produce the thawing type of initial condition, we require $n>0$. Since in our case $n=1$, we restrict our analysis to the thawing type of initial condition. So, in our analysis we consider thawing class of initial conditions to solve the autonomous system of equations in Eq. \eqref{eq:auto_sys}.

In the thawing class of initial conditions Galileon field has $ w_{\phi} \approx -1 $ initially. It is possible for $ x<<1 $ which we can see through first line of Eq. \eqref{eq:imp_bkg_qnt}; at $ x<<1 $, $ w_{\phi} \approx \frac{-12 y^{2}(\epsilon+1)}{3(4 \epsilon+4)y^{2}} = -1 $. So, we restrict ourselves as $ x_{i}  < 10^{-2} $. The subscript 'i' refers to the initial value of any quantity at $ z = 1100 $. Note that the evolution of the background quantities has no significant dependence on the $ x_{i} $ parameter as long as $ x_{i}<<1 $.

The initial condition in $ y $ can be transfered to the boundary condition in $ \Omega_{\phi} $ through second line of Eq. \eqref{eq:imp_bkg_qnt}. So, we find $ y_{i} $ by solving back $ \Omega_{\phi}^{(0)} = 0.7 $.

The initial value of $ \lambda $ controls the initial slopes of the potential. For $ \lambda_{i} << 1 $ the equation of state of the Galileon field does not deviate much from its initial value $ -1 $ i.e. it always stays very close to the cosmological constant behavior. For higher values of $ \lambda_{i} $, the Galileon field sufficiently thaws away from the cosmological constant behavior accordingly. So, in our analysis, we take a range of the $ \lambda_{i}$ parameter as $0 \leq \lambda_{i} \leq 1$ for all the models.

Finally, we vary the $ \epsilon_{i} $ parameter in the range given by $0 \leq \epsilon_{i} \leq 100$. So, we have three parameters ($ \epsilon_{i} $, $x_{i}$, and $\lambda_{i}$) in our case for the cubic Galileon model. Note that if we fix $ \epsilon_{i}=0 $, the value of $\epsilon$ remains zero for the entire background evolution. So, the canonical quintessence model with linear potential is represented here by $ c_{3} = \alpha = \beta = \epsilon_{i} = \epsilon = 0$.

%%%%%%%%%%%%%%%%%%%%%%%%%%%%%%%%%%
\begin{figure}[tbp]
\centering
\includegraphics[width=.7\textwidth]{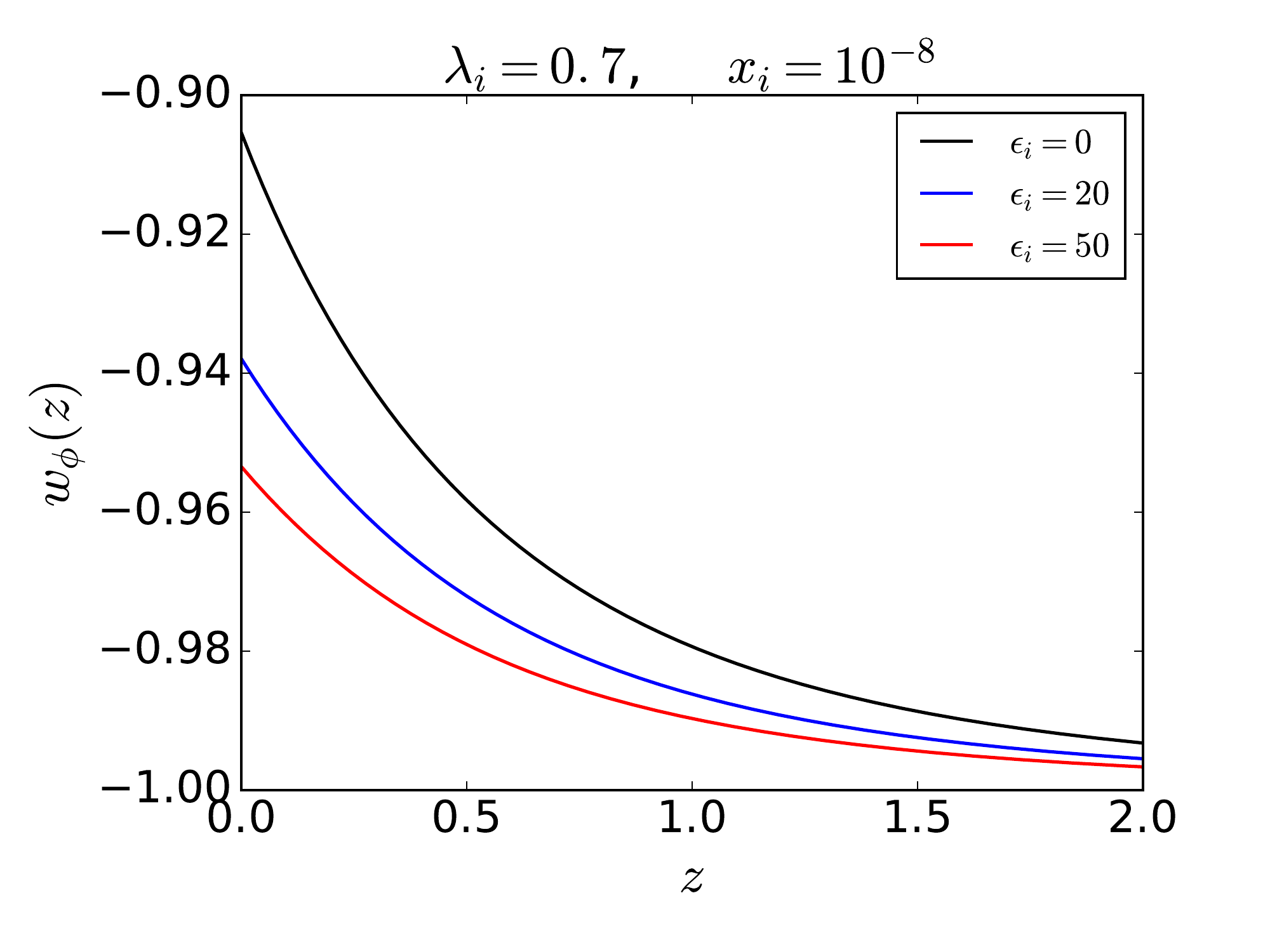}
\caption{\label{fig:eos} $ w_{\phi}(z) $ vs. $ z $ plots. The thawing type initial condition has been considered. So, $w_{\phi} \simeq -1$ at sufficiently early time. At late times, the value of $w_{\phi}$ increases from $-1$. Smaller the value of $\epsilon_{i}$ larger the deviation from the $\Lambda$CDM model.}
\end{figure}
%%%%%%%%%%%%%%%%%%%%%%%%%%%%%%%%%

\subsection{Behaviour of some background quantities}

Before proceeding to the study of perturbation or structure formation, let us discuss some important background quantities.

In Fig.~\ref{fig:eos} we have plotted equation of state of the Galileon field. At sufficiently early time all the Galileon models have $ w_{\phi} = -1 $ and as time increases $ w_{\phi} $ increases towards non-phantom values. $ \epsilon_{i} = 0 $ model has the highest deviation from the $ \Lambda $CDM model in the non-phantom side. The value of  $ w_{\phi} $ decreses with increasing $ \epsilon_{i} $. However, the value of $ w_{\phi} $ never goes to the phantom side. As the parameter $\epsilon_{i}$ increases, the cubic Galileon model gets closer to the $\Lambda$CDM model. But despite being increasing values of $\epsilon_{i}$, the cubic Galileon models remain non-phantom. So, the cubic Galileon model has always non-phantom behaviour.

%%%%%%%%%%%%%%%%%%%%%%%%%%%%%%%%%%
\begin{figure}[tbp]
\centering
\includegraphics[width=.495\textwidth]{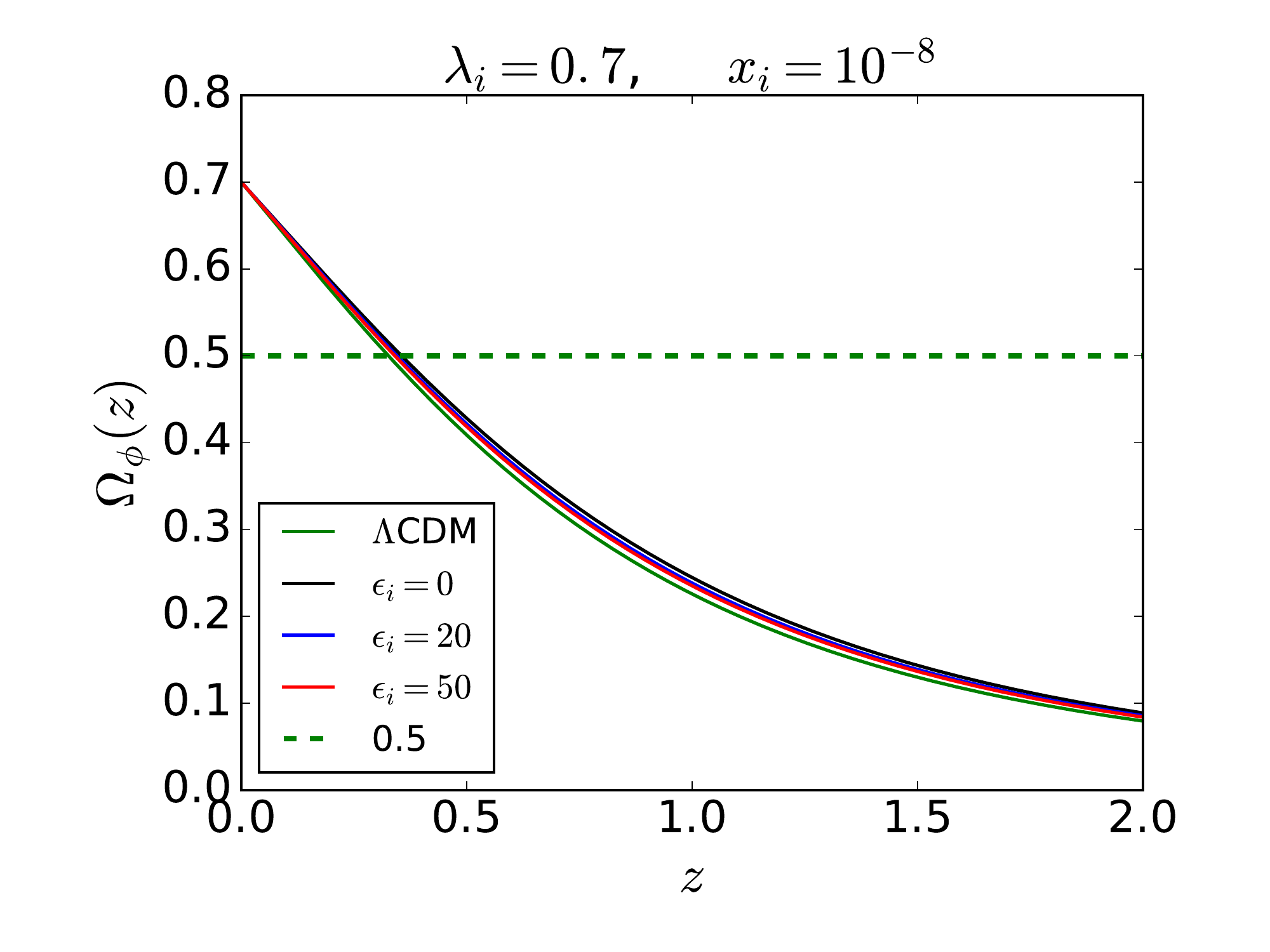}
\includegraphics[width=.495\textwidth]{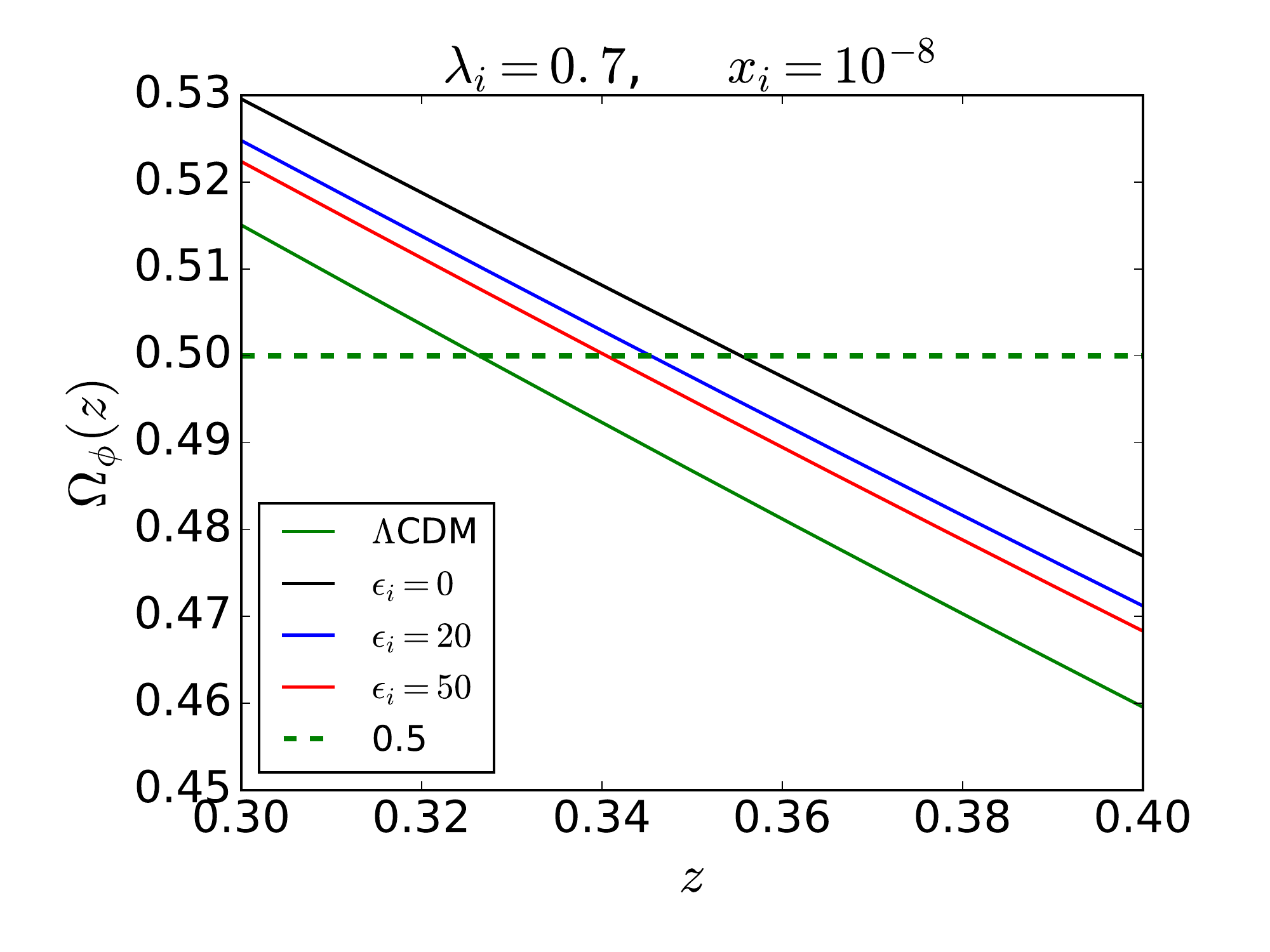}
\caption{\label{fig:Omegaq} $ \Omega_{\phi} $(z) vs. z plots. In all the models, dark energy dominates over the matter at redshifts $z\approx0.3-0.4$. Larger the non-phantom behavior (i.e. smaller the $\epsilon_{i}$ value) earlier the domination of the dark energy over the matter.}
\end{figure}
%%%%%%%%%%%%%%%%%%%%%%%%%%%%%%%%%%

In Fig.~\ref{fig:Omegaq} we have shown $ \Omega_{\phi} $(z) vs. z plots. The right panel is the zoomed in version of the left panel for the redshift range 0.3 to 0.4. The horizontal green-dashed line has been drawn at 0.5 to show at which redshift the Galileon field starts to dominate over total matter. In the $ \epsilon_{i} = 0 $ model, the Galileon field dominates earliest over the matter components. Larger the $ \epsilon_{i} $ value later the domination of the Galileon field over matter.

%%%%%%%%%%%%%%%%%%%%%%%%%%%%%%%%%%
\begin{figure}[tbp]
\centering
\includegraphics[width=.7\textwidth]{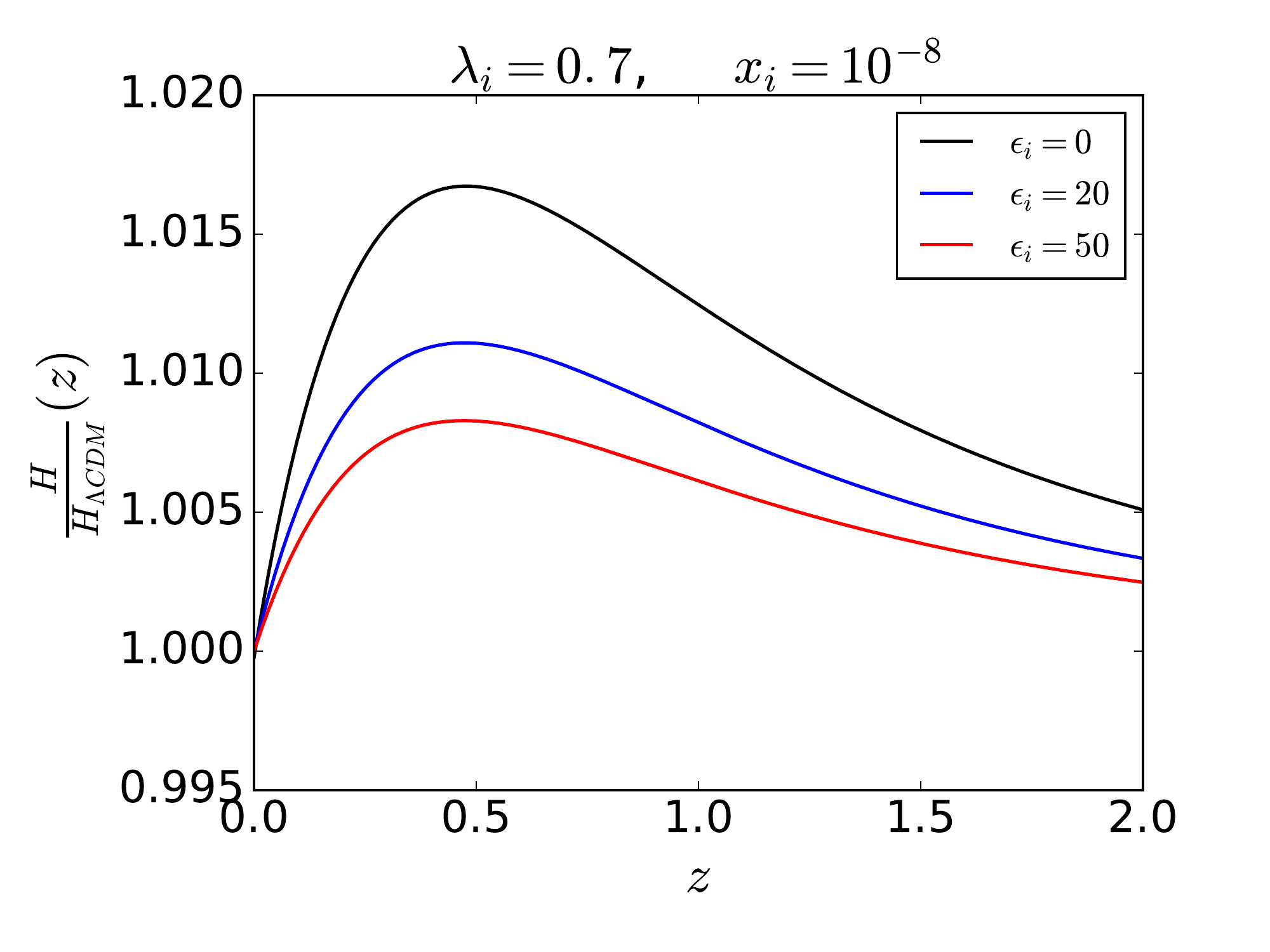}
\caption{\label{fig:Hubble} Deviations in the Hubble parameter compared to the $ \Lambda $CDM model. The value of the Hubble parameter increases with decreasing value of $\epsilon_{i}$ i.e. larger the non-phantom behavior larger the deviation in the Hubble parameter from the $\Lambda$CDM model.}
\end{figure}
%%%%%%%%%%%%%%%%%%%%%%%%%%%%%%%%%%

In Fig.~\ref{fig:Hubble}, we have plotted the deviations in the Hubble parameter from the $ \Lambda $CDM model. Earlier the domination of the Galileon field over matter corresponds to the earlier domination of the accelerated phase of the expansion of the Universe. So, expansion of the scale factor relatively increases if the Galileon field dominates earlier. Since smaller $ \epsilon_{i} $ value corresponds to the earlier domination of the Galileon field, the $ \epsilon_{i} = 0 $ model has the maximum positive deviation in the Hubble parameter from the $ \Lambda $CDM model. Larger the $ \epsilon_{i} $ value lower the deviation in the Hubble parameter. Since the cubic Galileon model always dominates earlier compared to the $ \Lambda $CDM model irrespective to the value of $ \epsilon_{i} $ (in the other way it is always non-phantom), the value of the Hubble parameter in cubic Galileon model is always higher than the $ \Lambda $CDM value.

%%%%%%%%%%%%%%%%%%%%%%%%%%%%%%%%%%
\begin{figure}[tbp]
\centering
\includegraphics[width=.495\textwidth]{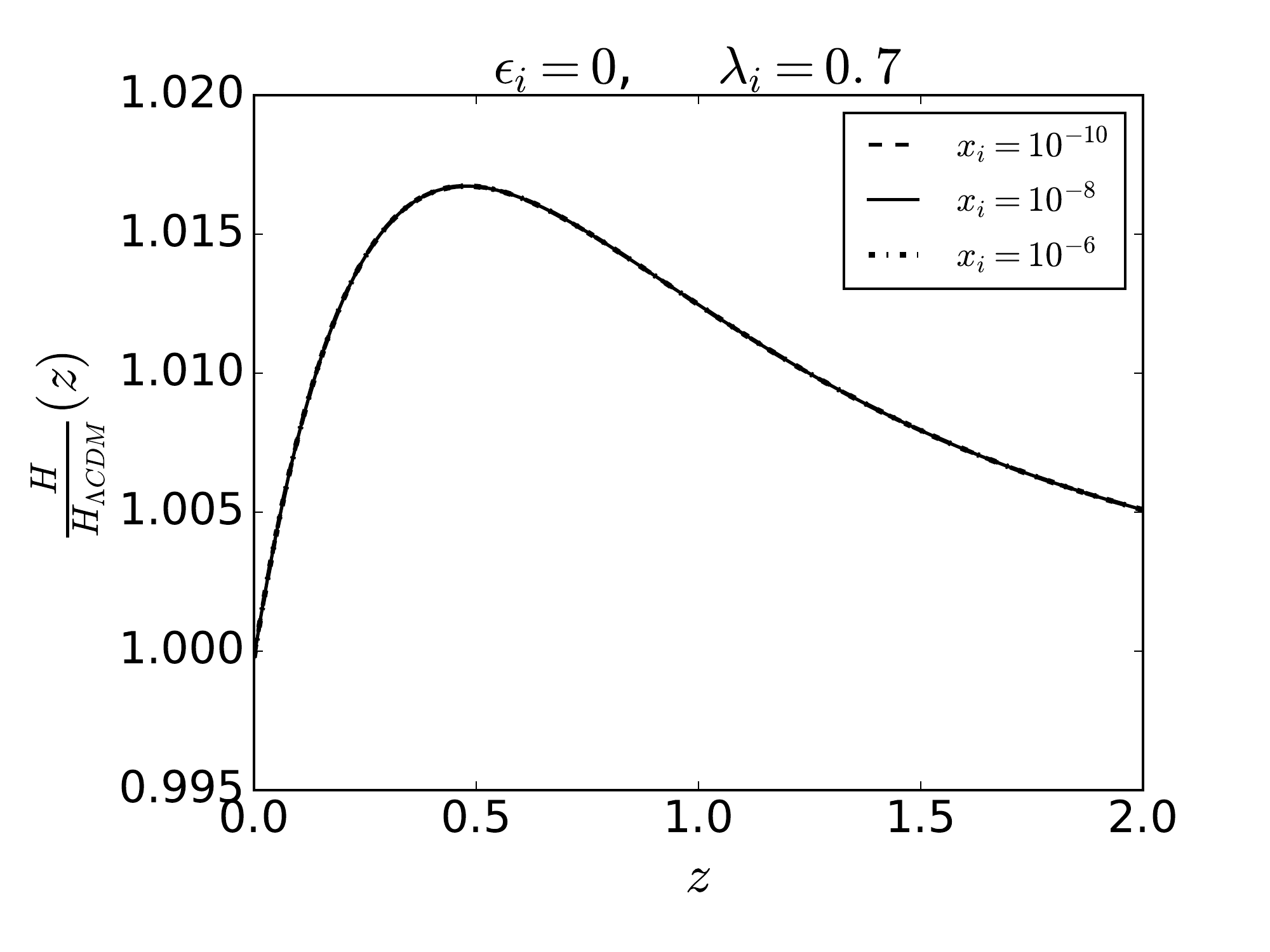}
\includegraphics[width=.495\textwidth]{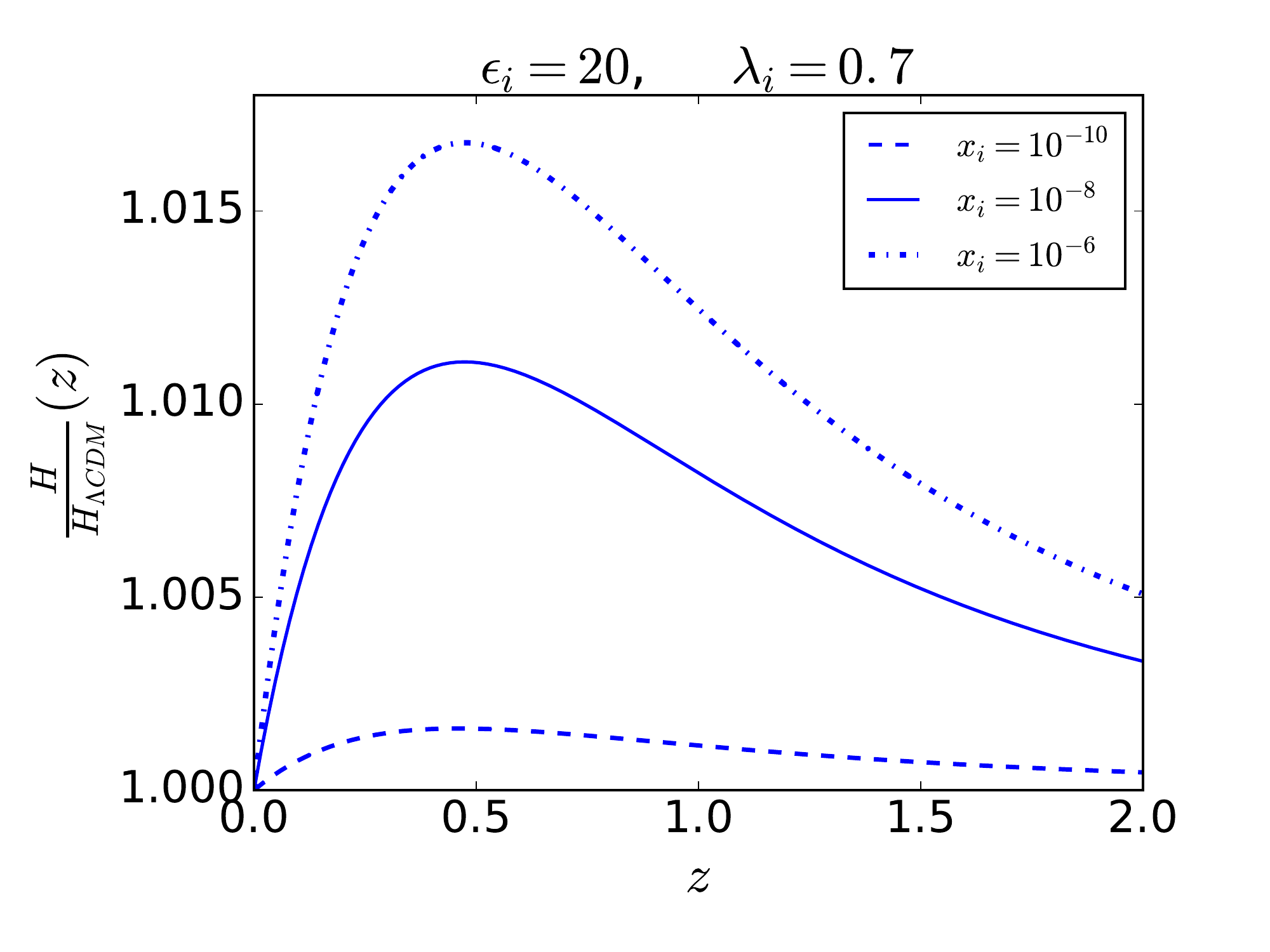}
\includegraphics[width=.495\textwidth]{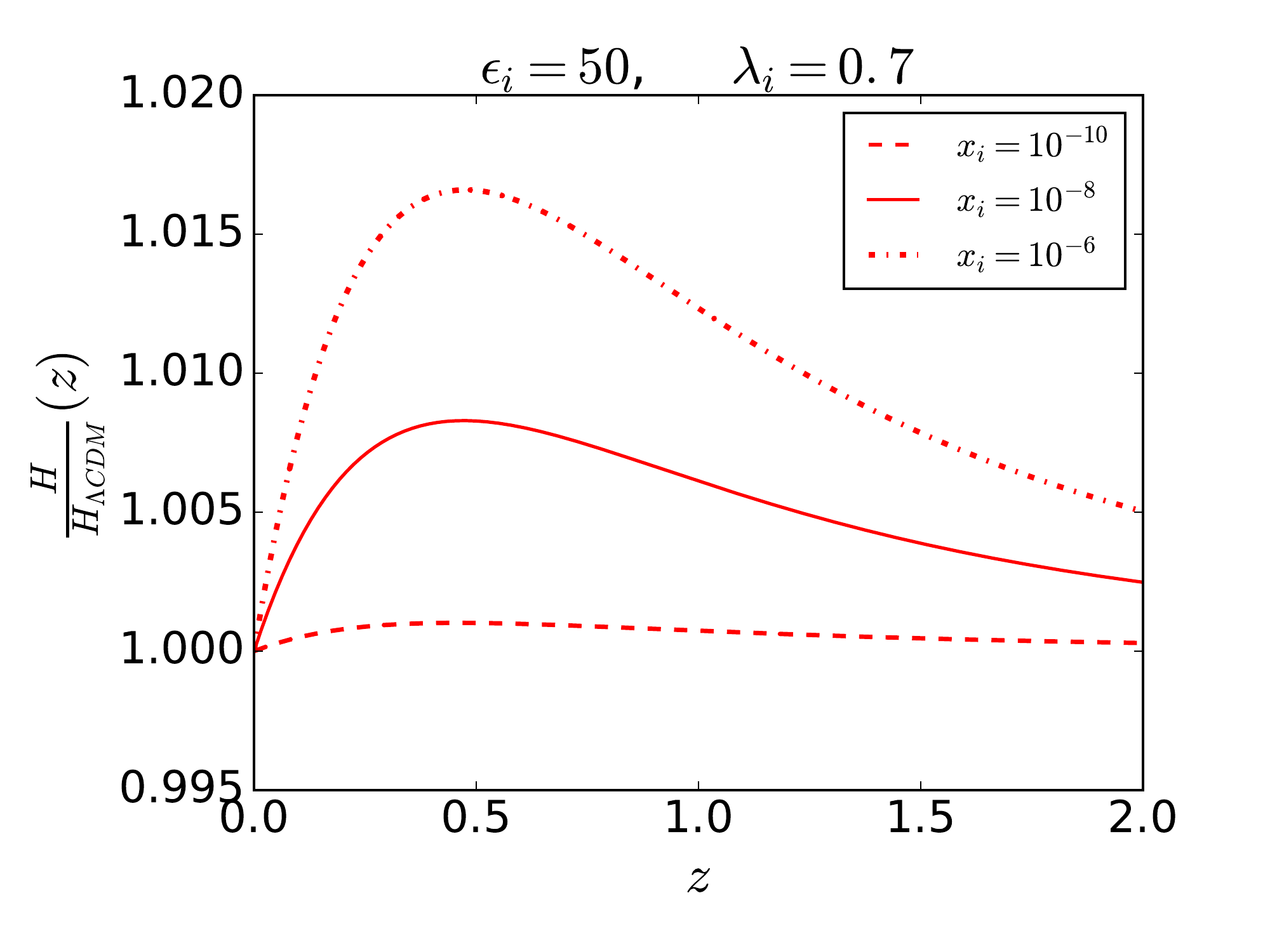}
\caption{\label{fig:Hubble2} Deviations in the Hubble parameter compared to the $ \Lambda $CDM model with the variation of the $x_{i}$ parameter. The top left panel shows that the dependency of the Hubble parameter on the $x_{i}$ parameter is negligible. However, for the cubic Galileon model (with $\epsilon_{i}>0$), the dependency is present but not much significant (at sub-percentage level).}
\end{figure}
%%%%%%%%%%%%%%%%%%%%%%%%%%%%%%%%%%

%%%%%%%%%%%%%%%%%%%%%%%%%%%%%%%%%%
\begin{figure}[tbp]
\centering
\includegraphics[width=.495\textwidth]{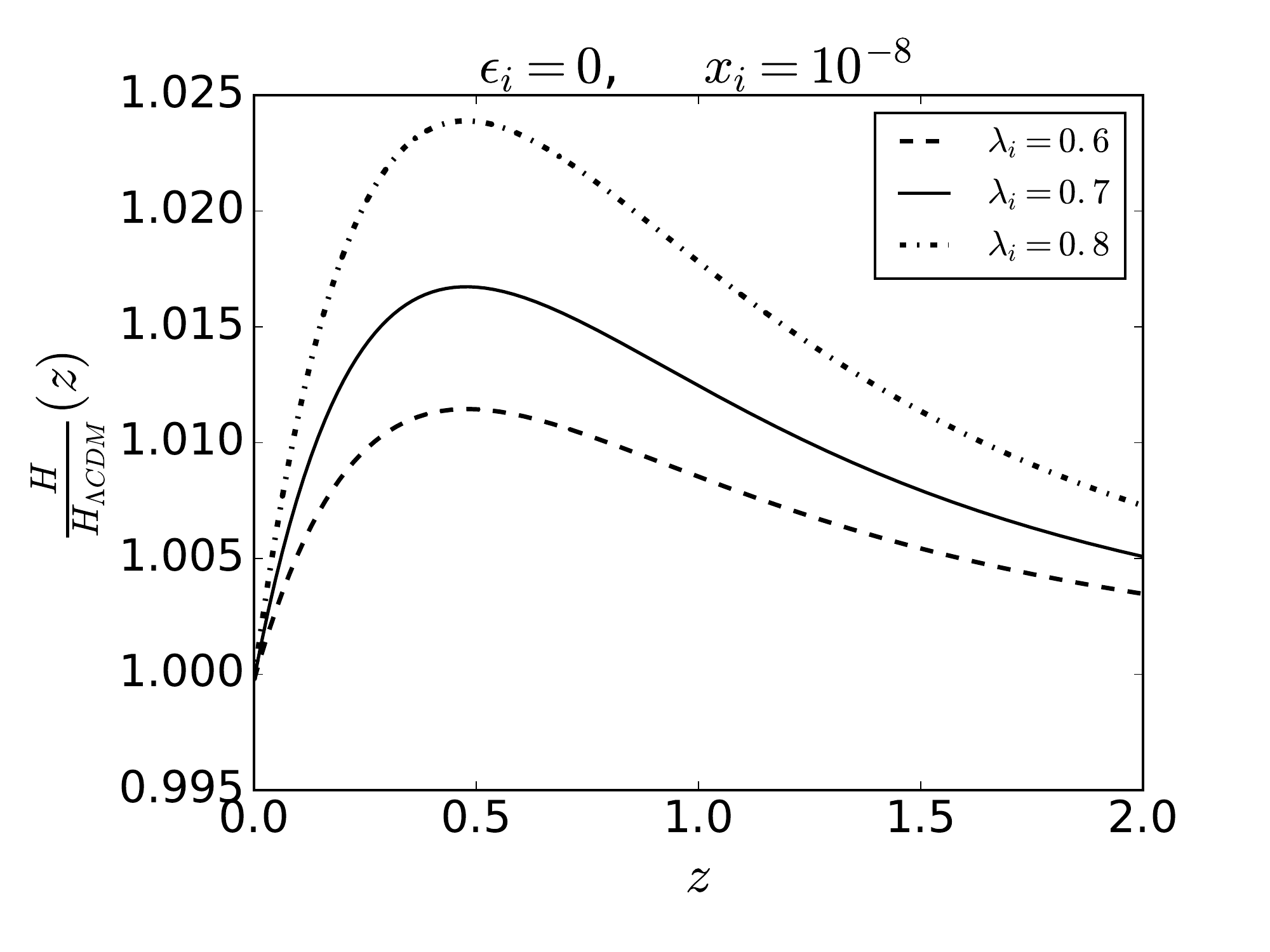}
\includegraphics[width=.495\textwidth]{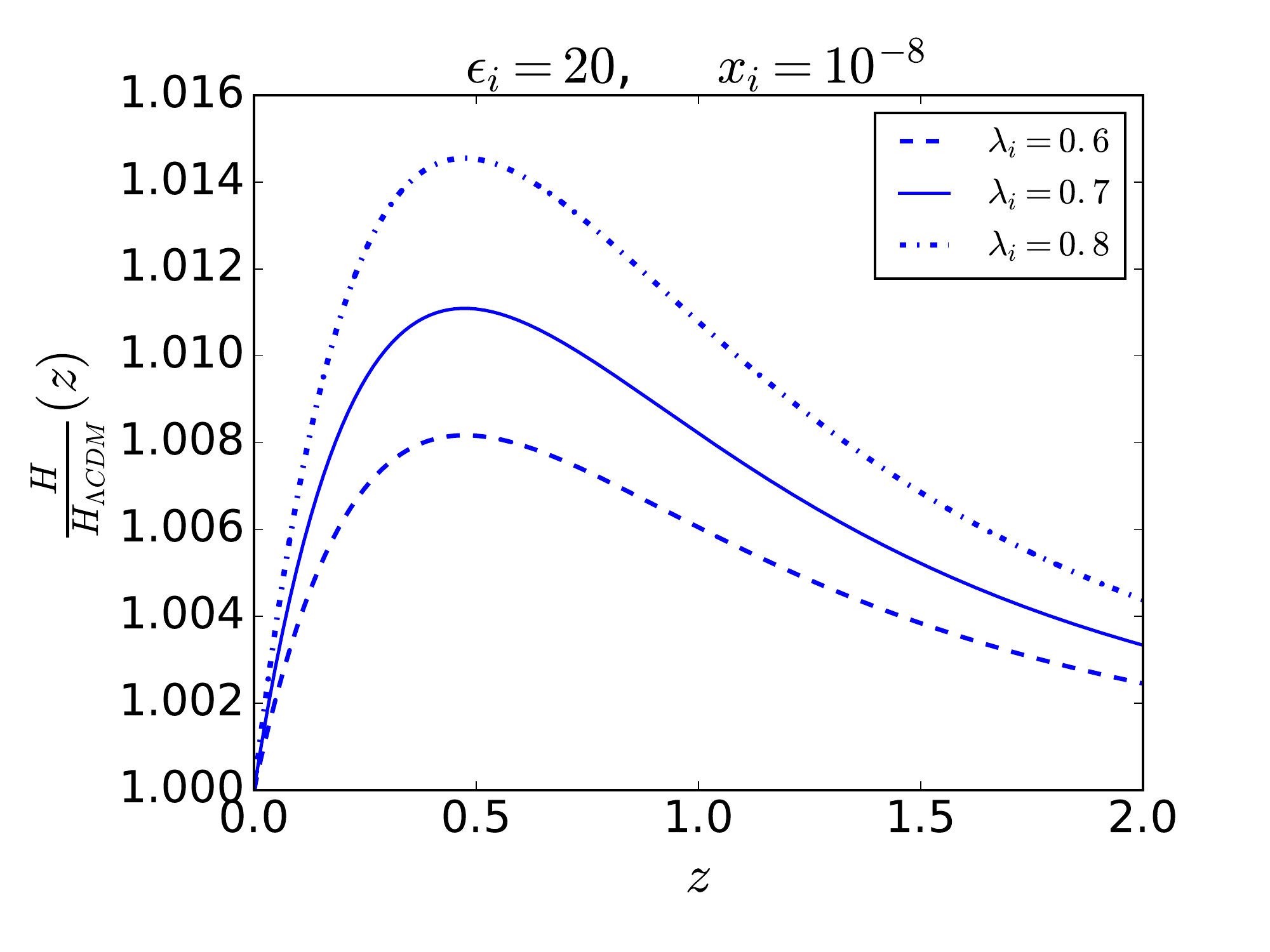}
\includegraphics[width=.495\textwidth]{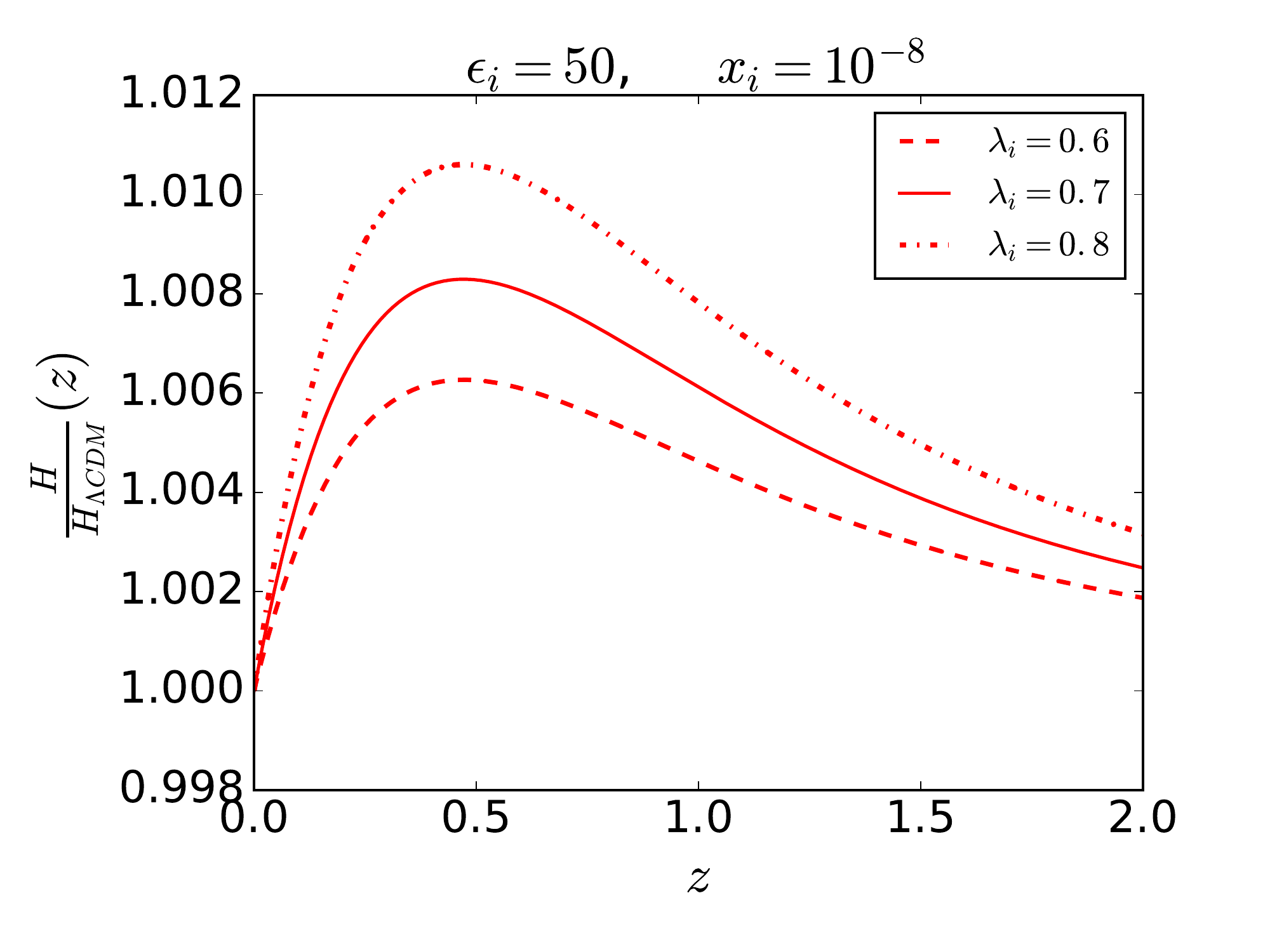}
\caption{\label{fig:Hubble3} Deviations in the Hubble parameter compared to the $ \Lambda $CDM model with the variation of the $\lambda_{i}$ parameter.  The deviations increase with increasing values of $\lambda_{i}$. The changes in the deviation are the largest in the quintessence model ($\epsilon_{i}=0$) and decrease with increasing values of $\epsilon_{i}$.}
\end{figure}
%%%%%%%%%%%%%%%%%%%%%%%%%%%%%%%%%%

%\noindent
So far we have analyzed the background dynamics by varying the $\epsilon_{i}$ parameter only. To see how the background quantities change with the changes of the $\lambda_{i}$ and $x_{i}$ parameters, we have plotted $H/H_{\Lambda CDM}$ graphs again in Figs.~\ref{fig:Hubble2} and ~\ref{fig:Hubble3}.

In Fig.~\ref{fig:Hubble2} we have varied $x_{i}$ keeping $\epsilon_{i}$ and $\lambda_{i}$ fixed. One interesting feature to see that for $x_{i} \ll 1$ the background dynamics for the quintessence model does not depend on $x_{i}$ significantly. For the positive values of $\epsilon_{i}$ in the cubic Galileon model, the background dynamics significantly depend on $x_{i}$. The Hubble parameter increases with the increasing $x_{i}$. However, these changes are not so significant for the cubic Galileon models too because the changes are at the sub-percentage level.

%\noindent
In Fig.~\ref{fig:Hubble3} we have shown the variation of the Hubble parameter with respect to the $\lambda_{i}$ values keeping $\epsilon_{i}$ and $x_{i}$ fixed. Both for quintessence and cubic Galileon models, the Hubble parameter increases with increasing values of $\lambda_{i}$. Another interesting point to notice that this increment is the highest for the quintessence model. The variation of the Hubble parameter with respect to $\lambda_{i}$ decrease with increasing values of $\epsilon_{i}$.

%\noindent
So, the dependency of the other background quantities on the $x_{i}$ and $\lambda_{i}$ parameters can be seen accordingly from the Figs.~\ref{fig:Hubble2} and ~\ref{fig:Hubble3} respectively (using the discussions above). To avoid many plots, next, we shall show the $x_{i}$ and $\lambda_{i}$ parameters dependency only in the final result i.e. during the discussions of the deviation in the convergence power spectrum. In the meantime, we restrict our discussions and the plots only by varying the $\epsilon_{i}$ parameter keeping $x_{i}=10^{-8}$ and $\lambda_{i}=0.7$ throughout.

\section{Evolution of perturbations}

In the quasi-static approximation (on sub-Hubble scale) to the full general relativistic perturbation, the linear evolution of the matter-energy density contrast has the standard form given by \citep{gal1,gal7}

\begin{equation}
\ddot{\delta}_{m}^{qs}+2 H \dot{\delta}_{m}^{qs} - 4 \pi G_{eff} \bar{\rho}_{m} \delta_{m}^{qs} = 0.
\label{eq:deltam_qs}
\end{equation}

\noindent
where $ G_{eff} = \left(1 + \frac{\beta ^2 \dot{\phi }^4}{2 A M_{pl}^2}\right) G = \left(1 + \frac{x^2 \epsilon ^2}{12 A}\right) G $ with $ G $ being the Newtonian gravitational constant and $ A $ is given by \citep{gal1,gal7}

\begin{eqnarray}
A &=& 1 -2 \beta  \left(2 H \dot{\phi }+ \ddot{\phi} \right)-\frac{\beta ^2 \dot{\phi }^4}{2 M_{pl}^2}
\nonumber\\
&=& \frac{x \left(-2 (B-4) \epsilon ^2+8 \epsilon +12\right)-x^3 \epsilon ^2 (\epsilon +1)+2 \sqrt{6} \lambda  y^2 \epsilon }{12 x (\epsilon +1)},
\end{eqnarray}

\noindent
where, $ B =3 + \frac{\dot{H}}{H^{2}} = 1.5(1-w_{\phi} \Omega_{\phi}) $. The superscript 'qs' to the matter-energy density constrast ($ \delta_{m} $) refers to the fact that the Eq. \eqref{eq:deltam_qs} is valid in the quasi-static approximation.

%%%%%%%%%%%%%%%%%%%%%%%%%%%%%%%%%%
\begin{figure}[tbp]
\centering
\includegraphics[width=.495\textwidth]{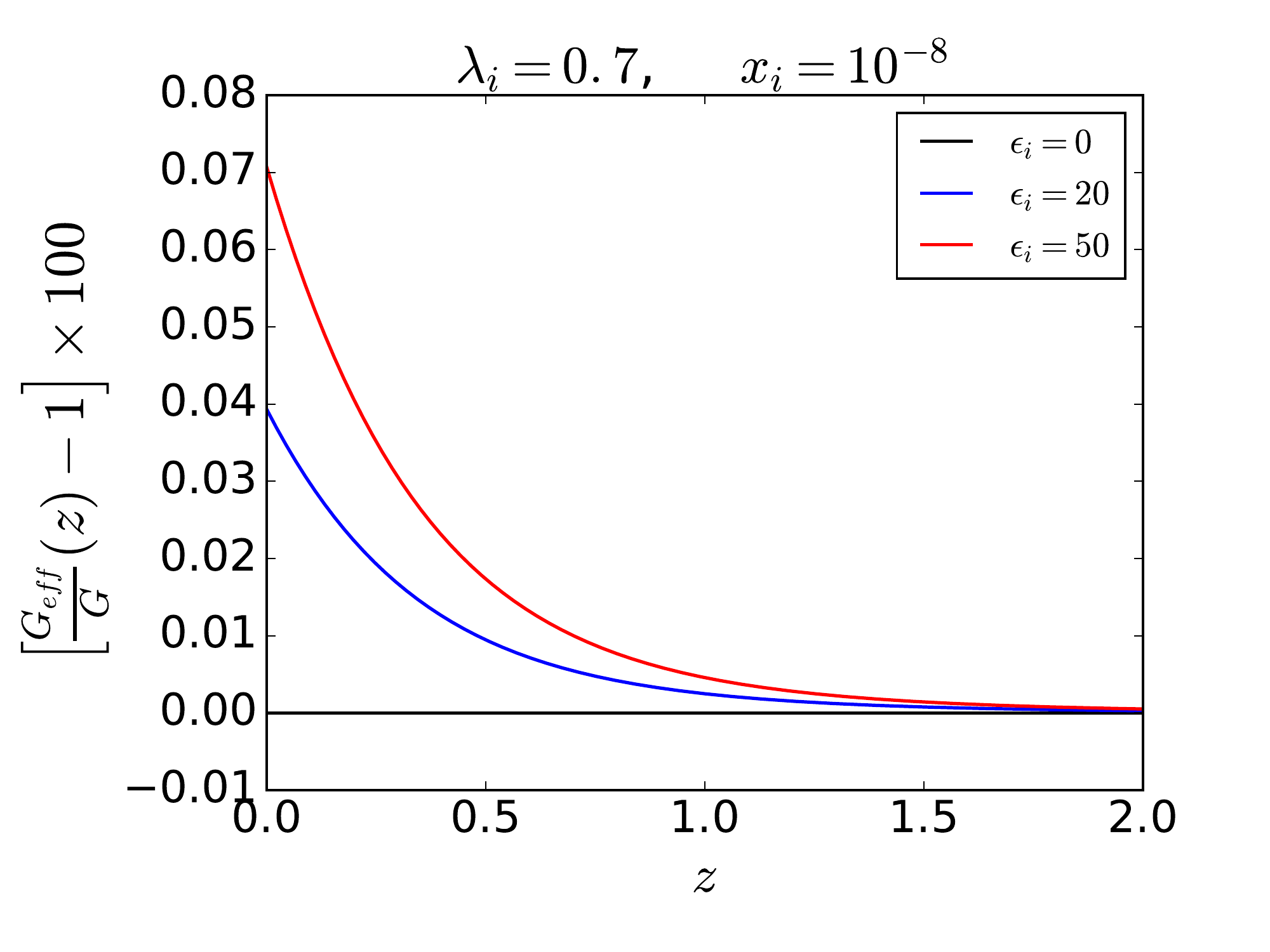}
\includegraphics[width=.495\textwidth]{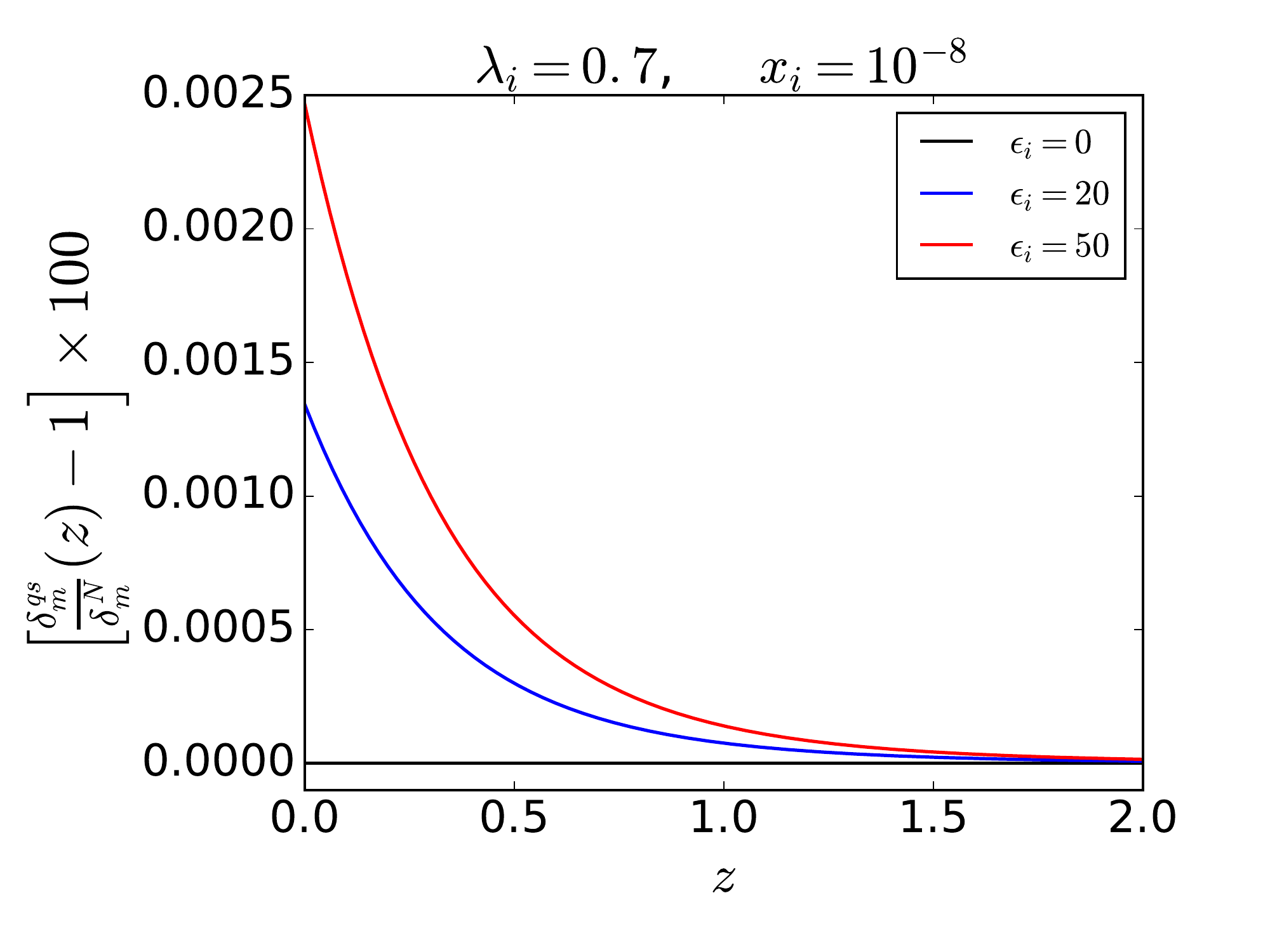}
\caption{\label{fig:qsvsN} Left panel: Percentage deviation of $ G_{eff} $ from $ G $. For the quintessence model ($\epsilon_{i}=0$) $G_{eff} = G$. For the cubic Galileon model, the deviation of the $G_{eff}$ from $G$ increases with increasing $\epsilon_{i}$. Right panel: Percentage deviation of $ \delta_{m}^{qs} $ from $ \delta_{m}^{N} $. Similarly, for the quintessence model $\delta_{m}^{qs} = \delta_{m}^{N}$. For the cubic Galileon model, the deviation of the $\delta_{m}^{qs}$ from $\delta_{m}^{N}$ increases with increasing $\epsilon_{i}$. However, both the deviations are highly insignificant (less than $0.1\%$).}
\end{figure}
%%%%%%%%%%%%%%%%%%%%%%%%%%%%%%%%%%

Since there is no coupling between matter and Galileon field, it is expected that $ G_{eff} \approx G $. To show that $ G_{eff} $ is indeed nearly equal to $ G $, in the left panel of Fig.~\ref{fig:qsvsN}, we have plotted percentage deviation of $ G_{eff} $ from $ G $. We can see that the deviations are less than $ 0.1 \% $. This result proves that in our case, we can simply use the Newtonian perturbation theory to study the perturbation in the sub-Hubble limit. In the Newtonian perturbation theory, the evolution equation of the matter-energy density contrast is given by \citep{LSSextra1,LSS1,LSS2,LSS3,LSS4,LSS5,LSS6,LSS7,LSS8,LSS9,LSSextra2}

\begin{equation}
\ddot{\delta}_{m}^{N}+2 H \dot{\delta}_{m}^{N} - 4 \pi G \bar{\rho}_{m} \delta_{m}^{N} = 0,
\label{eq:deltam_Nwtn}
\end{equation}

\noindent
where the superscript 'N' to the matter-energy density contrast ($ \delta_{m} $) refers to the fact that the Eq. \eqref{eq:deltam_Nwtn} is valid in the Newtonian approximation. To solve $ \delta_{m}^{qs} $ in Eq. \eqref{eq:deltam_qs} or $ \delta_{m}^{N} $ in Eq. \eqref{eq:deltam_Nwtn} we consider $ \delta_{m} \propto a $ for both initially at $ z = 1100 $; i.e. we consider $ \delta_{m}^{qs}|_{i} = \delta_{m}^{N}|_{i} = \frac{1}{1+1100} = \frac{d \delta_{m}^{qs}}{d N} \big{|}_{i} =\frac{d \delta_{m}^{N}}{d N} \big{|}_{i} $. Note that for both the $ \delta_{m} $ we consider growing mode solutions. In the right panel of Fig.~\ref{fig:qsvsN}, we have plotted percentage deviation of $ \delta_{m}^{qs} $ from $ \delta_{m}^{N} $ to show again that we can safely take Newtonian perturbation theory in our case. The deviations are less than $ 0.01 \% $. So, now onwards we shall use the Newtonian perturbation theory on the sub-Hubble scale and for simplicity, we omit the superscript 'N'.

%%%%%%%%%%%%%%%%%%%%%%%%%%%%%%%%%%
\begin{figure}[tbp]
\centering
\includegraphics[width=.7\textwidth]{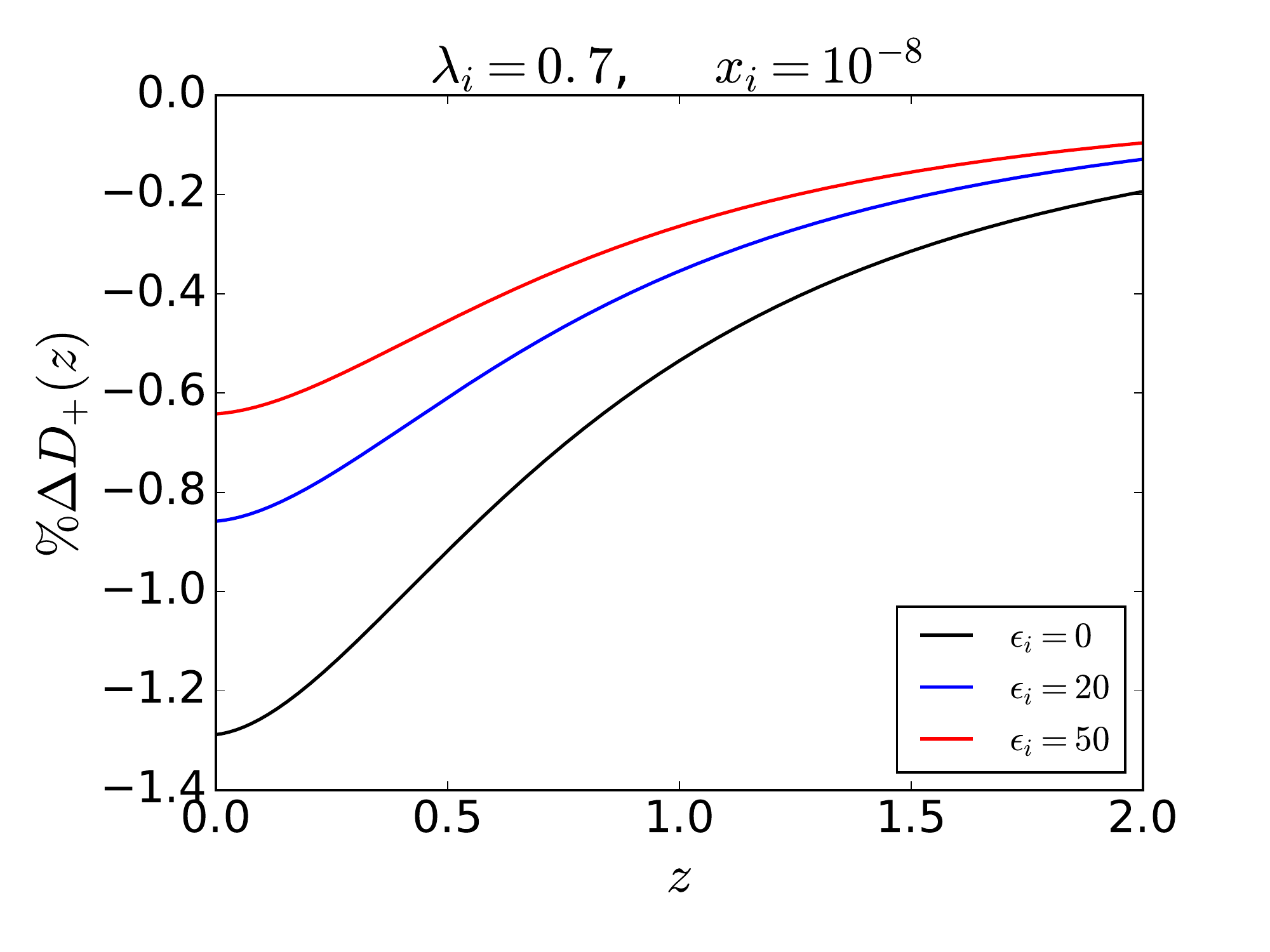}
\caption{\label{fig:growth} Percentage deviation in growing mode linear growth function from $ \Lambda $CDM model. Since in the non-phantom models, dark energy dominates over matter earlier i.e. the accelerated era starts earlier, the deviation in the growth function is negative for the non-phantom models compared to the $\Lambda$CDM model. Larger the non-phantom behavior (i.e. smaller the $\epsilon_{i}$ value) larger the negative deviation from the $\Lambda$CDM behavior.}
\end{figure}
%%%%%%%%%%%%%%%%%%%%%%%%%%%%%%%%%%

The growing mode solution can be described by a quantity called growth function ($ D_{+} $) which is defined as $ \delta_{m} (z) = D_{+} (z) \delta_{m}^{i} $. Rewriting Eq. \eqref{eq:deltam_Nwtn} with respect to $ N $ and using the definition of the growth function we get \citep{LSS2,LSSextra2}

\begin{equation}
\dfrac{d^{2} D_{+}}{d N^{2}} + \frac{1}{2} \Big{(} 1 - 3 w_{\phi} \Omega_{\phi} \Big{)} \dfrac{d D_{+}}{d N} - \frac{3}{2} \Omega_{m} D_{+} = 0
\label{eq:Dplus}
\end{equation}

\noindent
To solve the growing mode growth function we use the same initial conditions as in $ \delta_{m} $ i.e. $ D_{+} \propto a $ which gives $ D_{+}|_{i} = \frac{1}{1+1100} = \frac{d D_{+}}{d N} \big{|}_{i} $.

In Fig.~\ref{fig:growth} we have plotted the percentage deviations in the growing mode linear growth function for the cubic Galileon models compared to the $ \Lambda $CDM model. In this figure and in all the next figures, by the notation '$ \% \Delta $' we mean percentage deviation i.e. $ \% \Delta Q = \frac{Q_{G} - Q_{\Lambda CDM}}{Q_{\Lambda CDM}} \times 100 $ for any quantity, Q and subscript 'G' corresponds to it's value in the cubic Galileon model. During structure formation, the overdense regions grow with time due to the gravitational attraction between matters. The Galileon field has the negative pressure which produces repulsive gravitational force. The presence of the Galileon field slows down the expansion rate of the overdense regions due to this repulsive gravitational force. For this reason, the model in which Galileon field dominates earlier has the smaller growth rate. That is why in Fig.~\ref{fig:growth}, we can see that the deviations in $ D_{+} $ for the Galileon models are always negative compared to the $ \Lambda $CDM model as because cubic Galileon models are always non-phantom. Since in the $ \epsilon_{i} = 0 $ model, the Galileon field dominates earliest over matter, the deviations in the $ D_{+} $ for $ \epsilon_{i} = 0 $ model is the largest from the $ \Lambda $CDM model. Larger the $ \epsilon_{i} $ value smaller the deviation. So, the results in Fig.~\ref{fig:growth} are completely consistent with the results of the Figs.~\ref{fig:eos},~\ref{fig:Omegaq} and ~\ref{fig:Hubble}.

\subsection{Matter power spectrum}

The matter power spectrum, $ P_{m} $ is defined as $ < \tilde{\delta}_{m} (\vec{k},z) \hspace{0.1 cm} \tilde{\delta}_{m}^{*} (\vec{k'},z) > = (2 \pi)^{3} \delta^{3}_{D} (\vec{k}-\vec{k'}) P_{m} (k,z) $, where $ \tilde{\delta}_{m} $ is the fourier transform of the $ \delta_{m} $ and $ k $ is the magnitude of $ \vec{k} $. From this definition we find a relationship between the linear matter power spectrum and initial power spectrum at sufficiently early time given by \citep{LSSextra1,LSS2,LSSextra2}

\begin{equation}
P_{lin}(k,z) = \frac{D_{+}^{2}(z)}{D_{+}^{2}(z_{in})} P_{in}(k),
\label{eq:PSlinear}
\end{equation}

\noindent
where, $ P_{in}(k) = P_{lin}(k,z_{in}) $ is the initial matter power spectrum and for simplicity we have omitted the subscript 'm'. In our analysis, we have kept all the background and perturbation normalizations in a way that at a sufficiently early time (in an early matter dominated era) all the quantities remain same. So, for the power spectrum calculation, we have also taken same initial power spectrum for all the models (which corresponds to the same $A_{s}$ value for all the models). Note that although in principle at the early time (at $z=1100$) the power spectrum for all the models should differ, if we fix $A_{s}$, the difference is highly insignificant (much less than sub-percentage level). So, we can safely take the form of the power spectrum as in Eq. \eqref{eq:PSlinear}. We compute initial matter power spectrum from CAMB \citep{camb}. In the CAMB we first compute linear matter power spectrum at redshift $ z = 0 $ using $ \Lambda $CDM model for the cosmological parameters given by $ \Omega_{m}^{(0)} = 0.3 $, $ \Omega_{b}^{(0)} = 0.05 $, $ H_{0} = 100 h km/Sec/Mpc $ with $ h = 0.678 $, $ n_{s} = 0.968 $ and $ \sigma_{8}^{(0)} = \sigma_{8} (z = 0) = 0.83 $. Here, $ \Omega_{m}^{(0)} $, $ \Omega_{b}^{(0)} $ and $ H_{0} $ are the present day total matter energy density parameter, baryonic matter energy density parameter and Hubble parameter respectively. $ n_{s} $ be the scalar spectral index of the primordial perturbation. $ \sigma_{8}^{(0)} $ is the present day value of the rms fluctuation of mass within the boxes of $ 8 h^{-1} Mpc $ length scale. With these cosmological parameter values in the $ \Lambda $CDM model the value $ \sigma_{8}^{(0)} = 0.83 $ corresponds to the scalar power spectrum amplitude $ A_{s} \simeq 2.27 \times 10^{-9} $ at pivot scale $ k_{p} = 0.05 Mpc^{-1} $ related to the primordial curvature perturbation. These values are consistant with the Plank 2015 results \citep{Planck,Planck2}. After computing linear matter power spectrum at redshift $ z = 0 $, we use linear matter growth function for $ \Lambda $CDM model to compute the matter power spectrum at initial time ($ z_{in} = 1100 $). We use this initial matter power spectrum for all the models i.e. we consider same initial matter power spectrum for all the models.

%%%%%%%%%%%%%%%%%%%%%%%%%%%%%%%%%%
\begin{figure}[tbp]
\centering
\includegraphics[width=.495\textwidth]{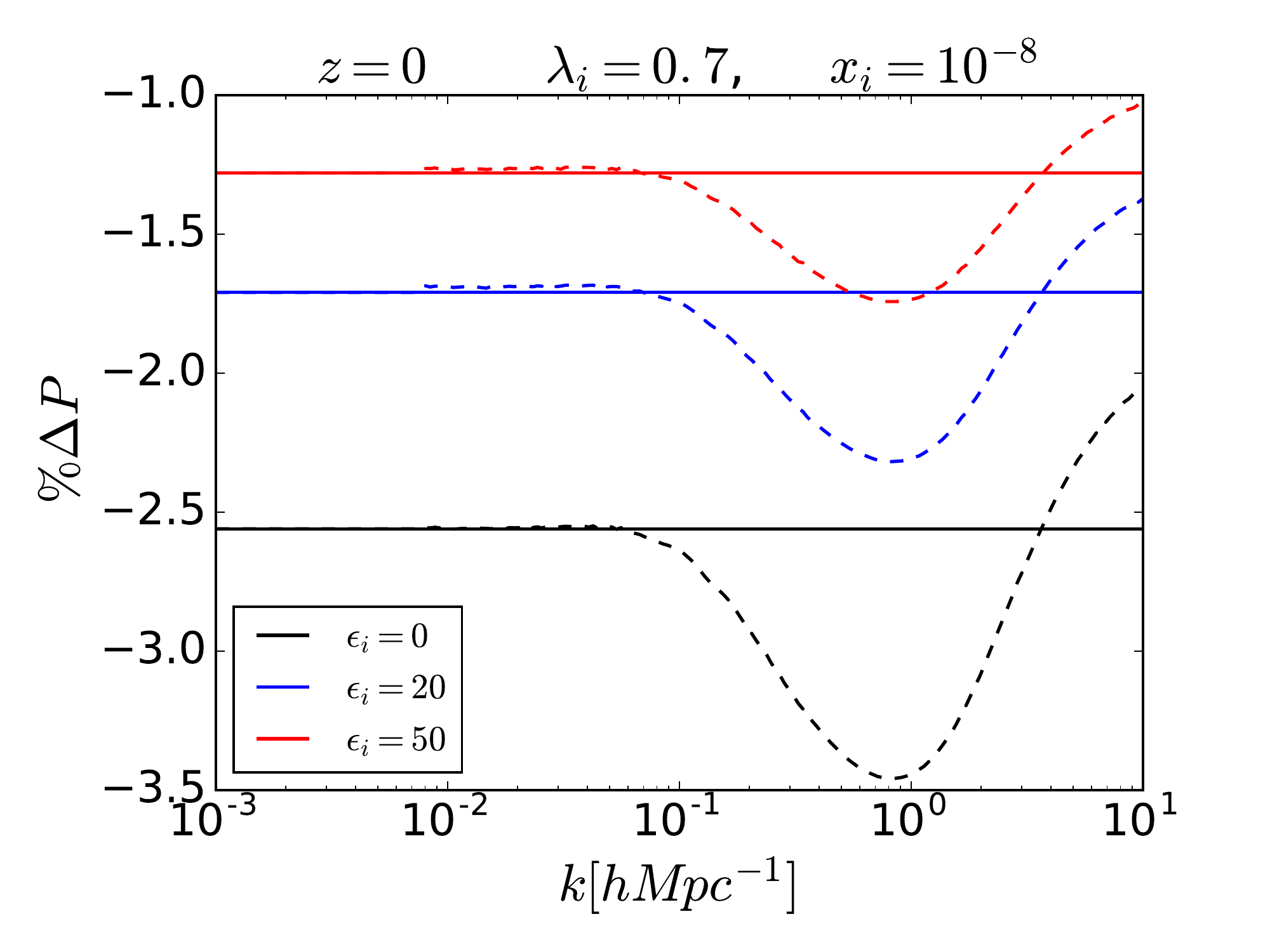}
\includegraphics[width=.495\textwidth]{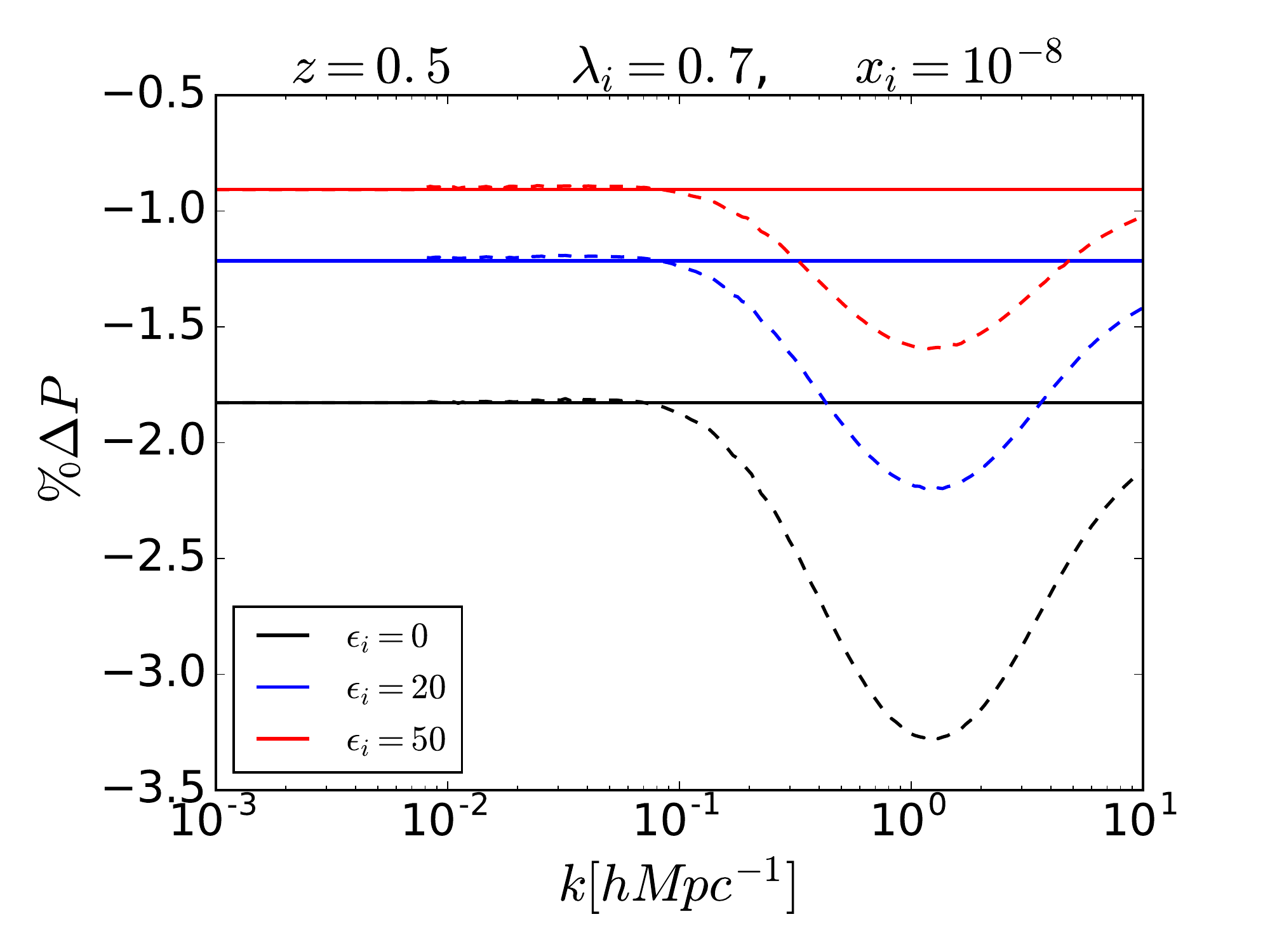}\\
\includegraphics[width=.495\textwidth]{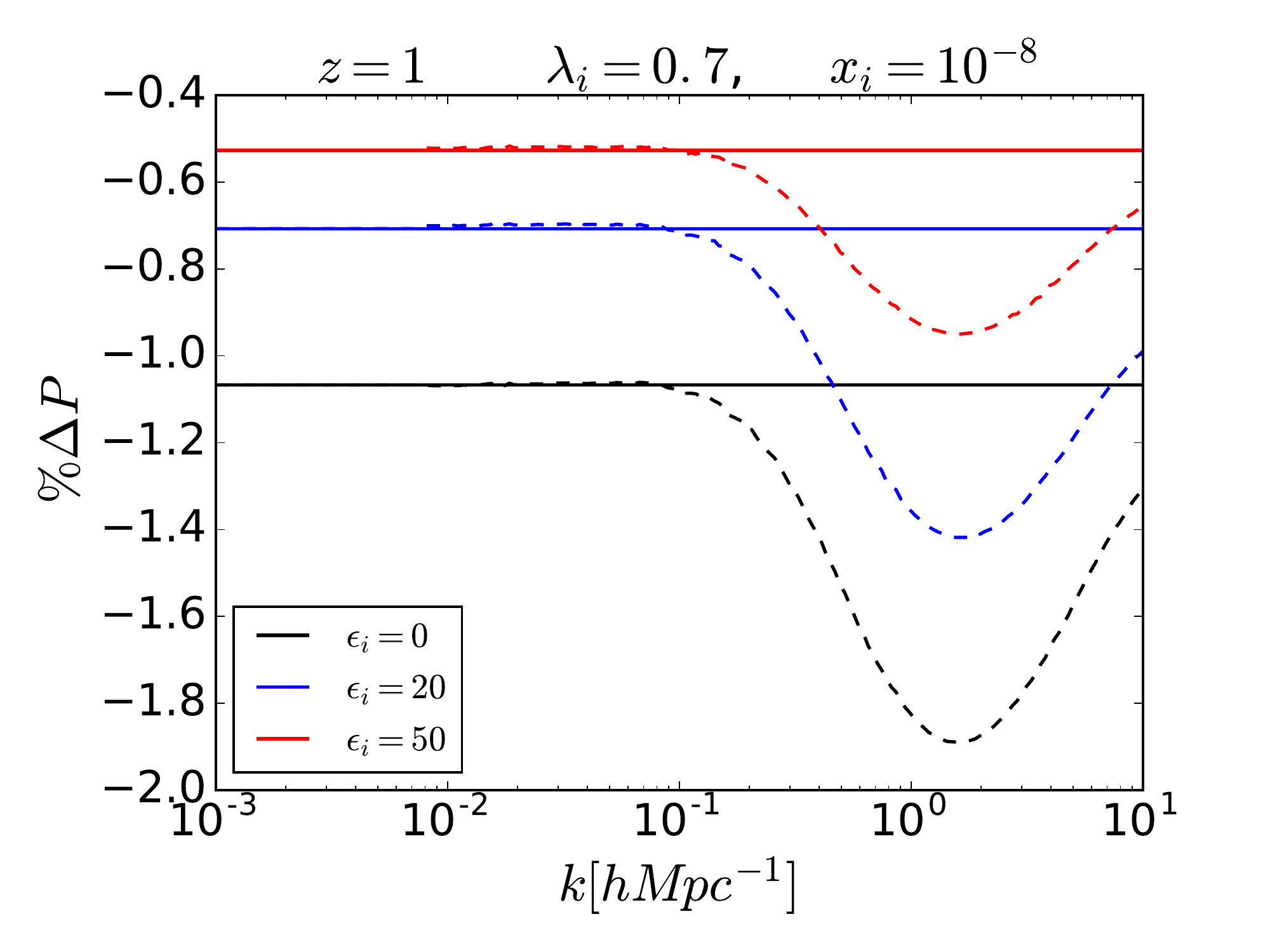}
\includegraphics[width=.495\textwidth]{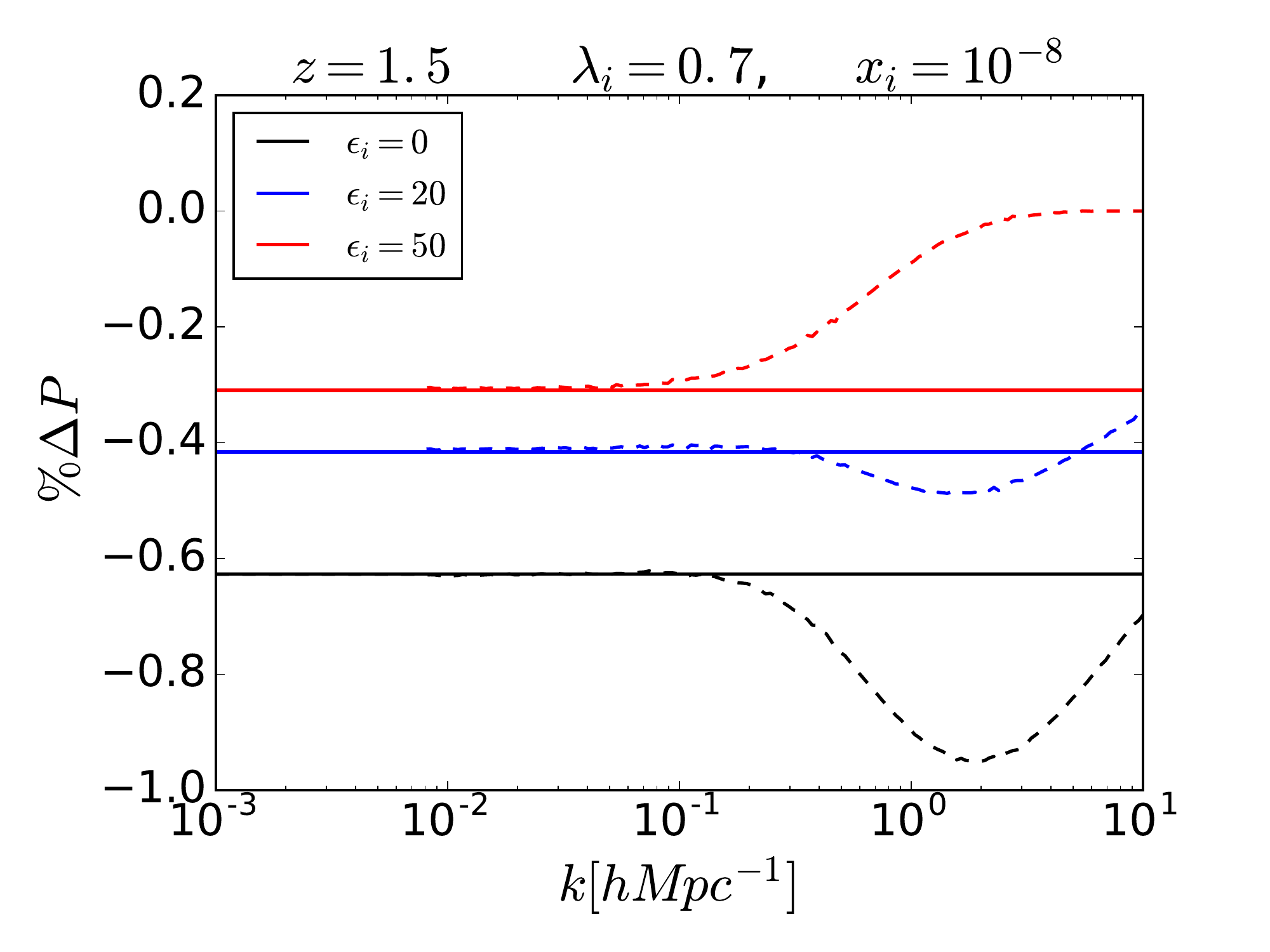}
\caption{\label{fig:ps} Percentage deviation in the matter power spectrum from the $ \Lambda $CDM model. Continuous lines are for the linear matter power spectrum described in Eq. \eqref{eq:PSlinear}. Dashed lines are for the non-linear matter power spectrum from HMCode \citep{HMCode}. The behavior is consistent as in Fig.~\ref{fig:growth} at each redshift. The only difference is that the amplitude of the deviation in the matter power spectrum is roughly the twice the deviation in the growth function at each redshift accordingly.}
\end{figure}
%%%%%%%%%%%%%%%%%%%%%%%%%%%%%%%%%%

In Fig.~\ref{fig:ps}, we have plotted percentage deviation in the matter power spectrum for cubic Galileon model with different $ \epsilon_{i} $ values compared to the $ \Lambda $CDM model. The continuous lines are for the deviations in the linear matter power spectrum using Eq. \eqref{eq:PSlinear}. The dashed lines correspond to the deviations in the non-linear matter power spectrum. To compute the non-linear matter power spectrum we use HMCode (\citep{HMCode}) for the halofit ppf module in CAMB. The purpose to include the deviations in the non-linear matter power spectrum is that although in the non-linear regime the absolute value of the non-linear matter power spectrum deviates from the linear matter power spectrum, the deviations between two models remains similar (up to a percent only) both in the linear and non-linear matter power spectrum. The deviations in the matter power spectrum are always negative due to the same reason that Cubic Galileon model is always non-phantom irrespective of the value of $ \epsilon_{i} $. The $ \epsilon_{i} = 0 $ model has the highest deviation from the $ \Lambda $CDM model. Larger the value of $ \epsilon_{i} $ smaller the deviations. Since linear matter power spectrum is proportional to the square of the linear matter growth functions, the deviations in the linear matter power spectrum are roughly twice the deviations in the linear matter growth functions.

In this figure and in all the next figures (except for the Figs.~\ref{fig:wlps2} and ~\ref{fig:wlps3}) dashed lines correspond to the results involving non-linear matter power spectrum (computed by HMCode) replacing the linear matter power spectrum. The same color corresponds to the same model with continuous and dashed lines correspond to the results using linear and non-linear matter power spectrum respectively.

\subsection{Matter bi-spectrum}

The matter bi-spectrum, $ B $ is defined as $ < \tilde{\delta}_{m} (\vec{k}_{1},z) \hspace{0.1 cm} \tilde{\delta}_{m} (\vec{k}_{2},z) \hspace{0.1 cm} \tilde{\delta}_{m} (\vec{k}_{3},z) > = (2 \pi)^{3} \delta^{3}_{D} (\vec{k}_{1}+\vec{k}_{2}+\vec{k}_{3}) B(\vec{k}_{1}, \vec{k}_{2}, \vec{k}_{3}; z) $. In the Newtonian approximation the tree-level matter bi-spetrum is given by \citep{LSS2,LSSextra2}

\begin{equation}
B_{tree}(\vec{k}_{1}, \vec{k}_{2}, \vec{k}_{3}; z) = 2 F_{2}(\vec{k}_{1}, \vec{k}_{2}, z) P_{lin} (k_{1}, z) P_{lin} (k_{2}, z) + \text{2-cycles},
\label{eq:BStree}
\end{equation}

\noindent
where with the Einstein De-Sitter (EDS) approximation the $ F_{2} $ is given by \citep{LSS2,LSSextra2}

\begin{equation}
F_{2}(\vec{k}_{i},\vec{k}_{j}) \simeq \frac{5}{7} + \frac{1}{2} \Big{(} \frac{k_{i}}{k_{j}} + \frac{k_{j}}{k_{i}} \Big{)} \hat{k}_{i}.\hat{k}_{j} + \frac{2}{7} (\hat{k}_{i}.\hat{k}_{j})^{2},
\label{eq:F2solnEDSappx}
\end{equation}

Due to the different mode couplings in different triangular shapes, the tree-level matter bi-spectrum is accurate on the scale where $ k < 0.1 h Mpc^{-1} $. To increase the accuracy of the matter bi-spectrum in weakly non-linear or non-linear regime, we need high precision N-body simulations. However, to get a rough idea about the matter bi-spectrum on non-linear scales, we follow \citep{LSS10} (which is the improved version of \citep{LSS11}). According to \citep{LSS10}, the effective matter bi-spectrum which is valid on both linear and non-linear scales given by

\begin{equation}
B_{eff}(\vec{k}_{1}, \vec{k}_{2}, \vec{k}_{3}; z) = 2 F_{2}^{eff}(\vec{k}_{1}, \vec{k}_{2}, z) P_{NL} (k_{1}, z) P_{NL} (k_{2}, z) + \text{2-cycles},
\label{eq:BSeff}
\end{equation}

\noindent
where $ P_{NL} $ is the non-linear matter power spectrum and $ F_{2}^{eff} $ is given by \citep{LSS10}

\begin{eqnarray}
F_{2}^{eff}(\vec{k}_{i},\vec{k}_{j}) = \frac{5}{7} \tilde{a}(n_{i},k_{i}) \tilde{a}(n_{j},k_{j}) +&& \frac{1}{2} \Big{(} \frac{k_{i}}{k_{j}} + \frac{k_{j}}{k_{i}} \Big{)} (\hat{k}_{i}.\hat{k}_{j}) \tilde{b}(n_{i},k_{i}) \tilde{b}(n_{j},k_{j}) \nonumber\\
&& + \frac{2}{7} (\hat{k}_{i}.\hat{k}_{j})^{2}  \tilde{c}(n_{i},k_{i}) \tilde{c}(n_{j},k_{j}),
\label{eq:F2eff}
\end{eqnarray}

%%%%%%%%%%%%%%%%%%%%%%%%%%%%%%%%%%
\begin{figure}[tbp]
\centering
\includegraphics[width=.495\textwidth]{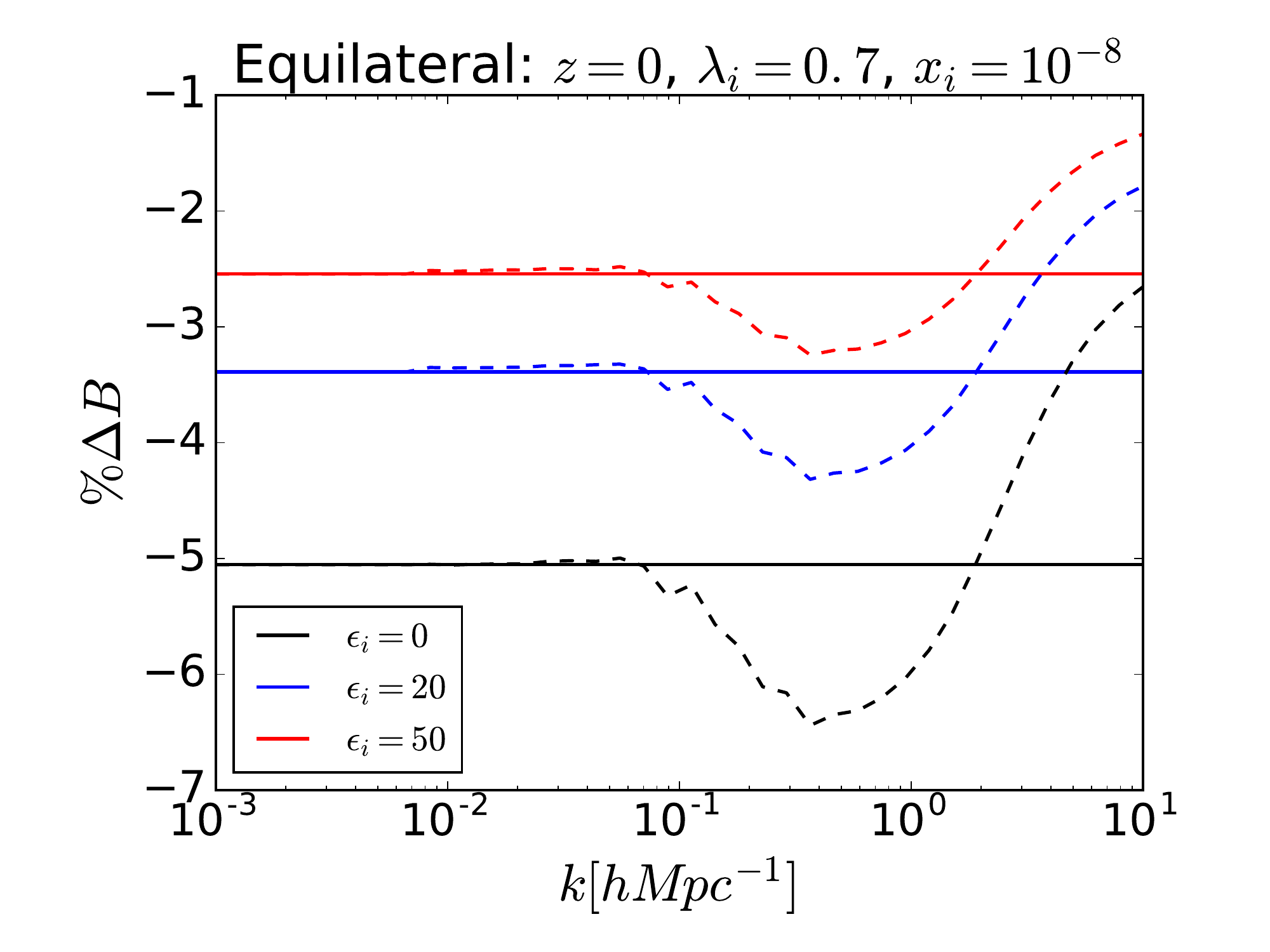}
\includegraphics[width=.495\textwidth]{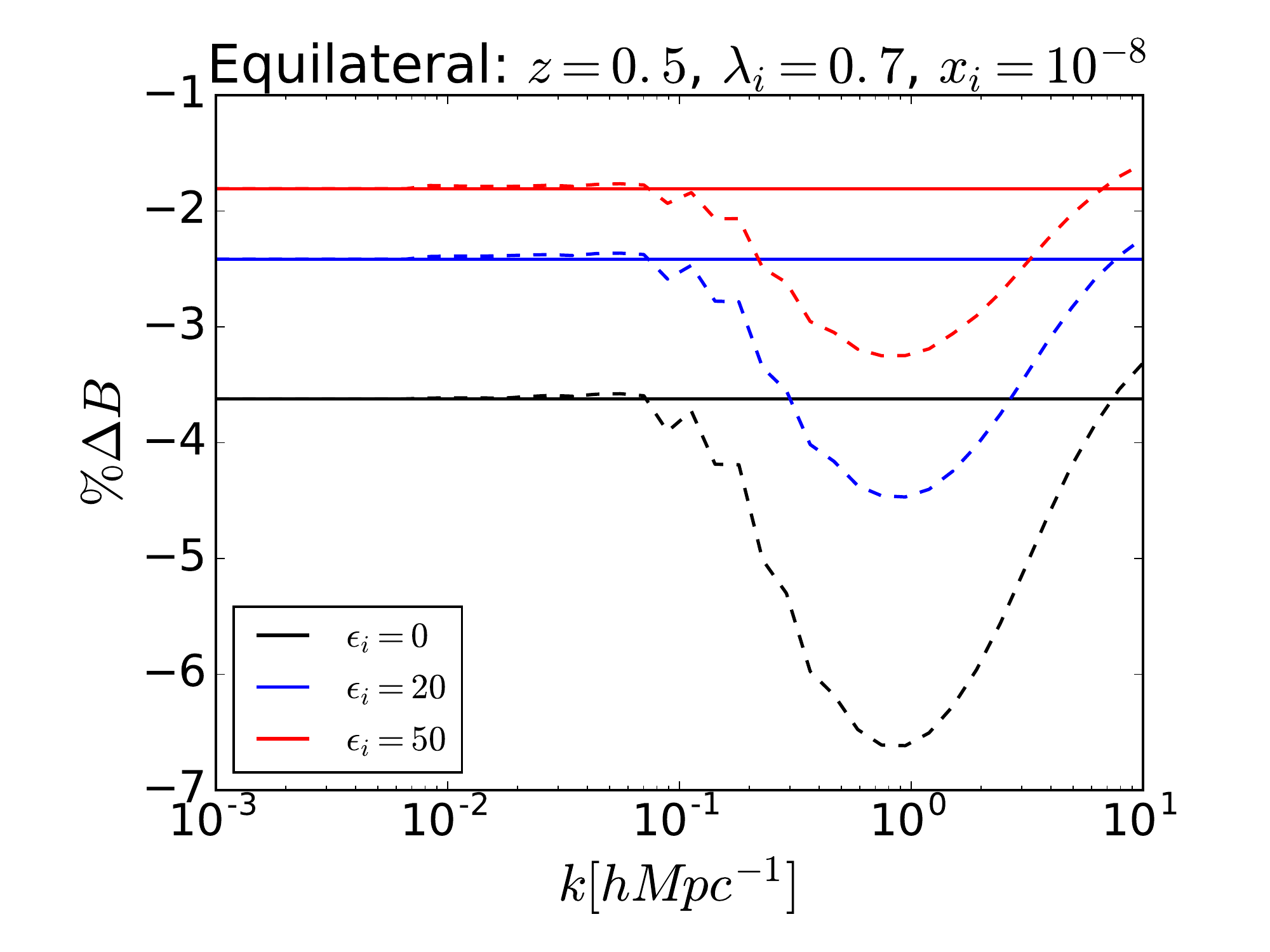}\\
\includegraphics[width=.495\textwidth]{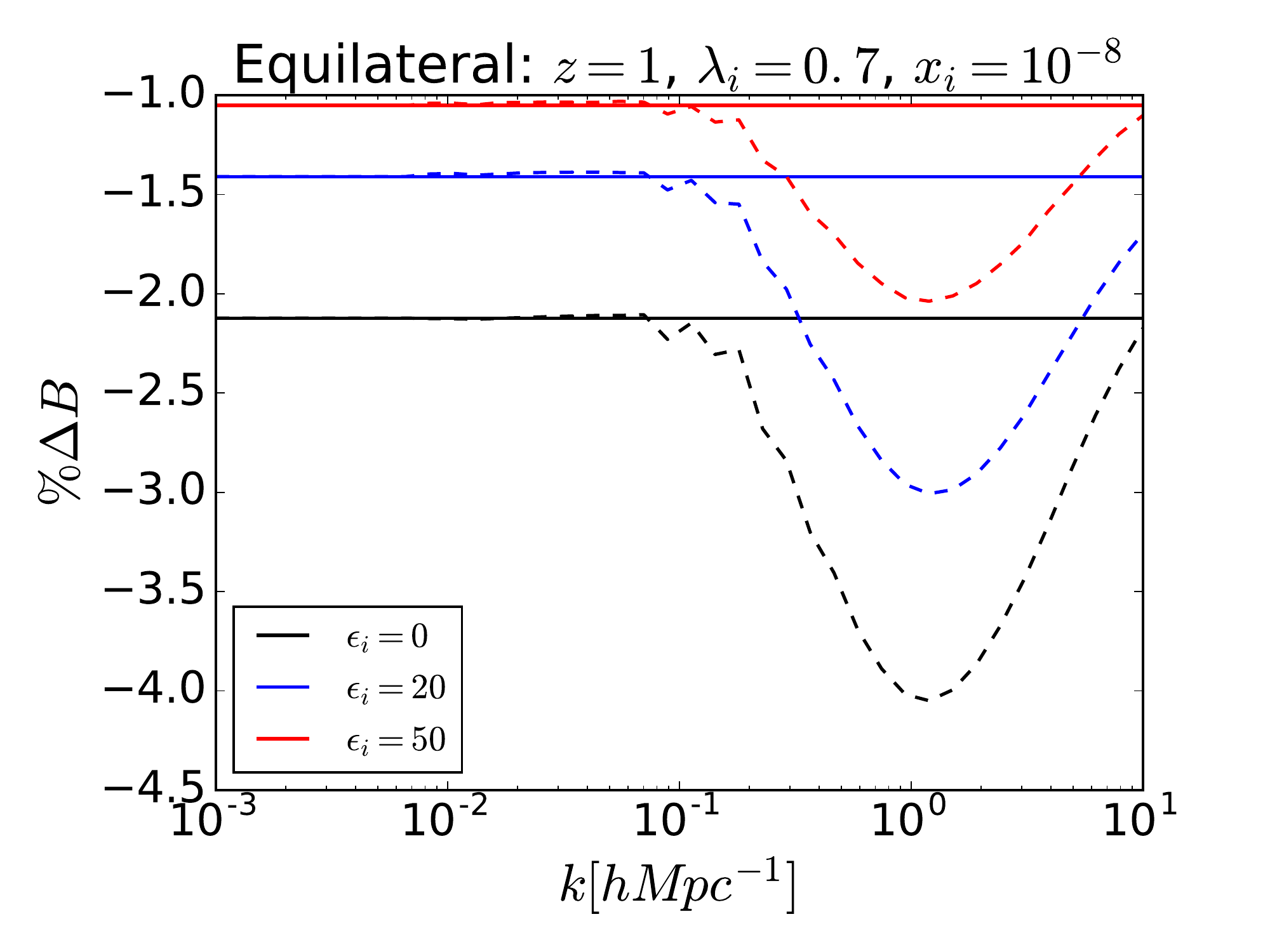}
\includegraphics[width=.495\textwidth]{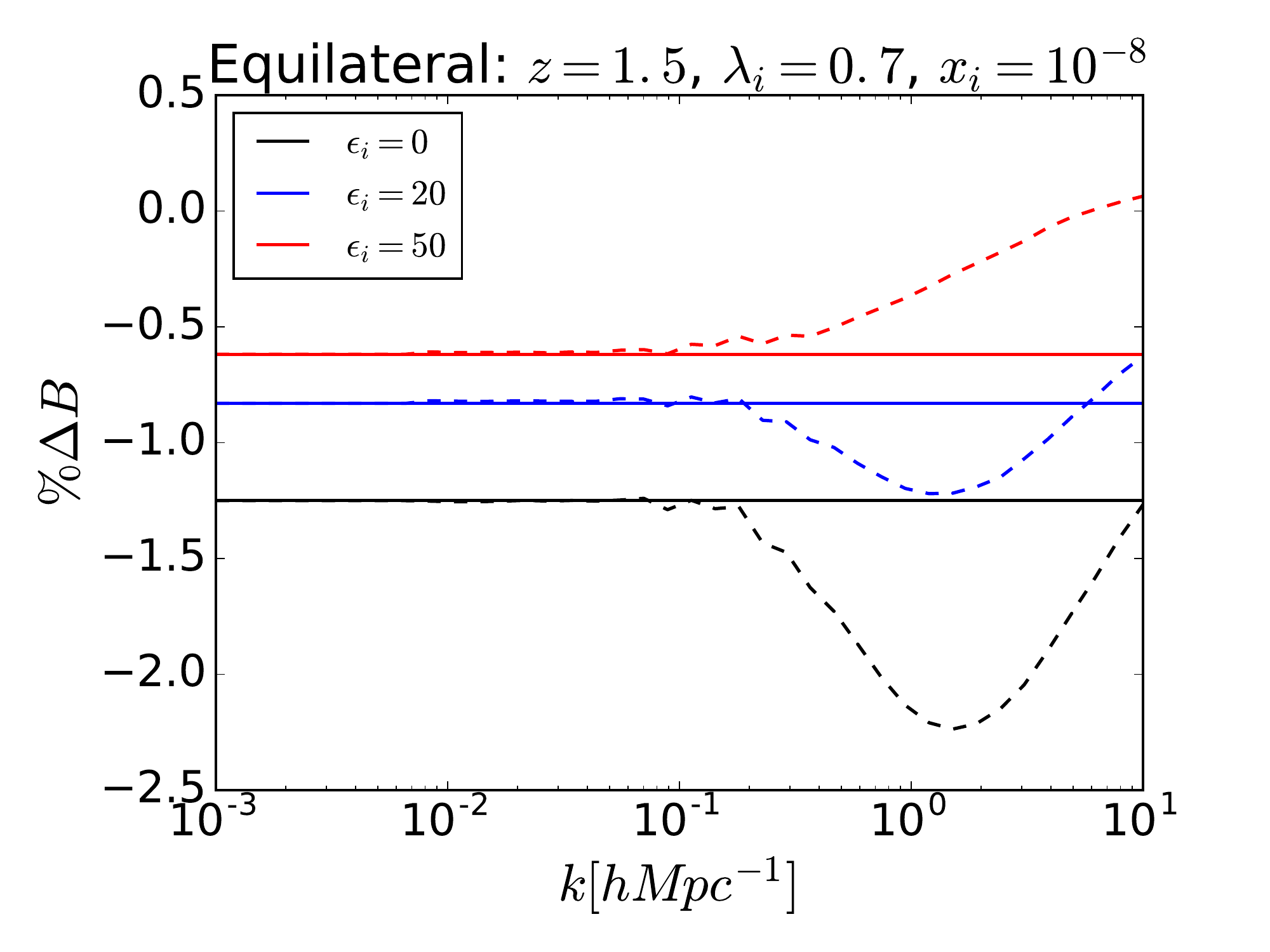}
\caption{\label{fig:bsequilateral} Percentage deviation in the matter bi-spectrum from $ \Lambda $CDM model for the equilateral configuration. Continuous lines are for the tree-level matter bi-spectrum described in Eq. \eqref{eq:BStree}. Dashed lines are for the non-linear effective matter bi-spectrum described in Eq. \eqref{eq:BSeff}. The behavior is consistent as in Fig.~\ref{fig:ps}. The only difference is that the amplitude of the deviation in the matter bispectrum is roughly the twice the deviation in the matter power spectrum accordingly.}
\end{figure}
%%%%%%%%%%%%%%%%%%%%%%%%%%%%%%%%%%

%%%%%%%%%%%%%%%%%%%%%%%%%%%%%%%%%%
\begin{figure}[tbp]
\centering
\includegraphics[width=.495\textwidth]{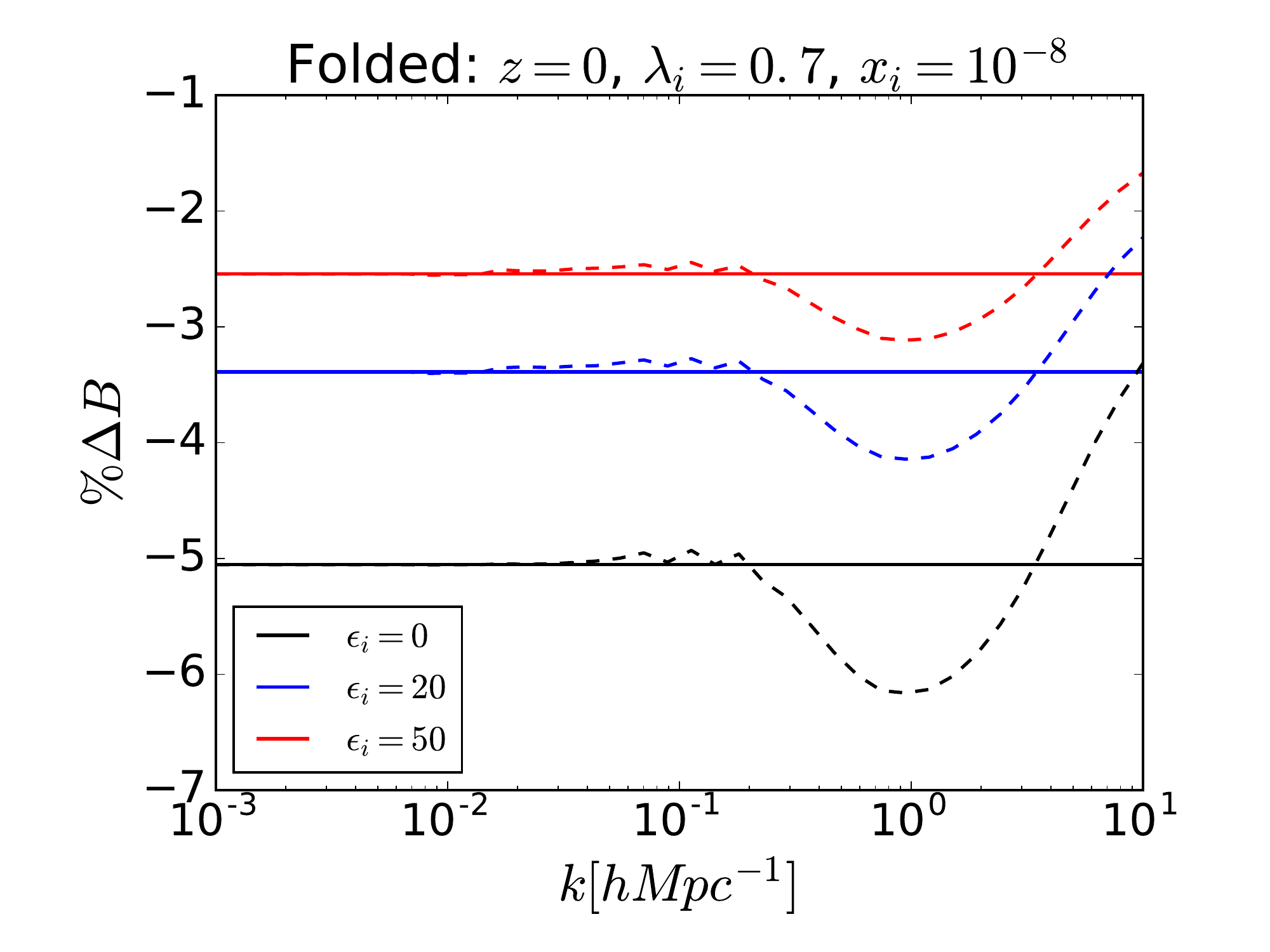}
\includegraphics[width=.495\textwidth]{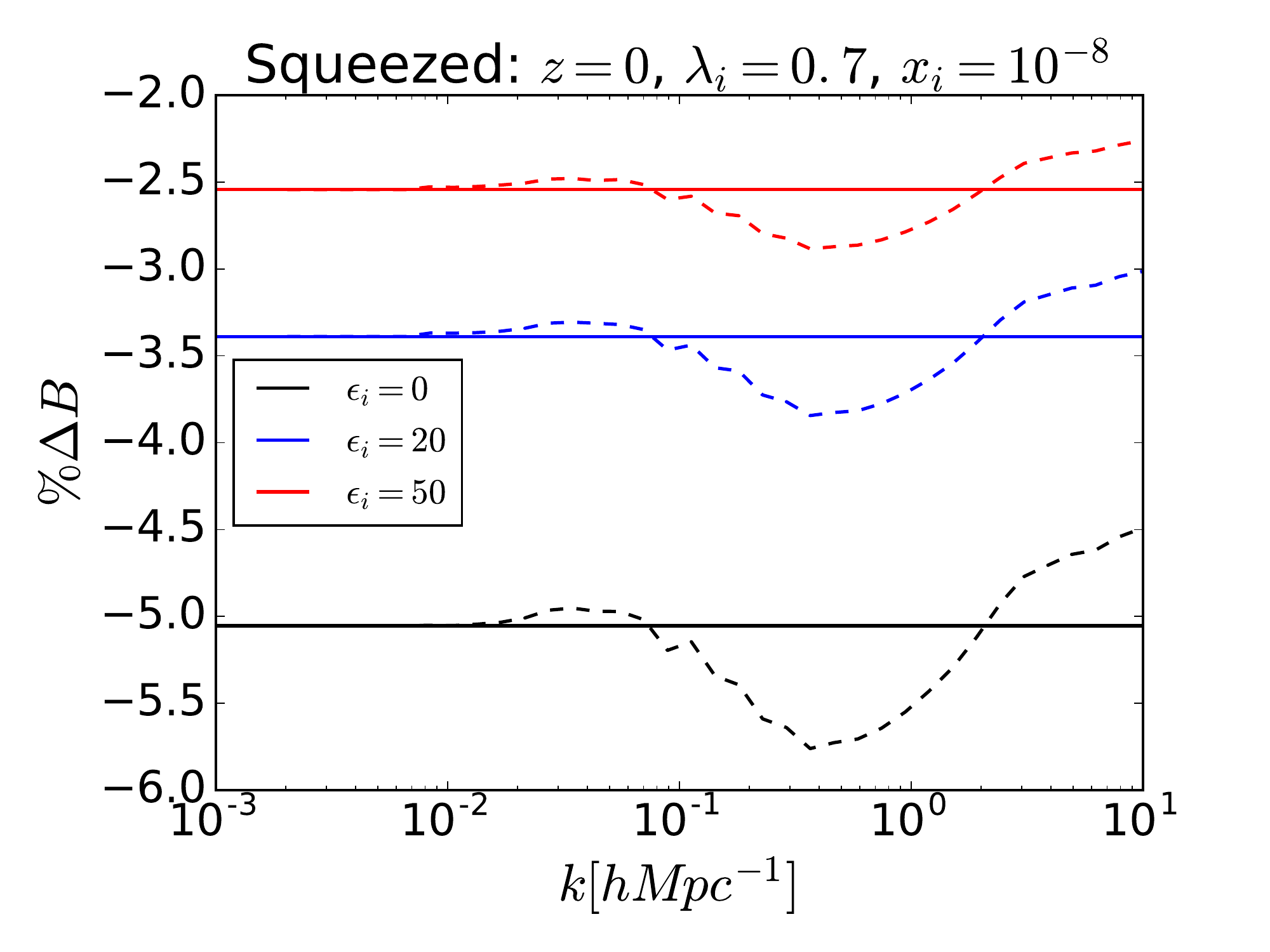}
\caption{\label{fig:bsFldpSqzd} Plots similar to the top-left panel of the Fig.~\ref{fig:bsequilateral} but for the folded (left panel) and squeezed (right panel) configurations respectively. The deviations are almost similar in the three configurations which can be seen by comparing this figure with the top left panel of Fig.~\ref{fig:bsequilateral}.}
\end{figure}
%%%%%%%%%%%%%%%%%%%%%%%%%%%%%%%%%%

\noindent
where we can see $ F_{2} $ in Eq. \eqref{eq:F2solnEDSappx} gets modified by $ \tilde{a} $, $ \tilde{b} $ and $ \tilde{c} $ factors which are given by \citep{LSS10}

\begin{eqnarray}
\tilde{a}(n,k) &=& \frac{1+\sigma_{8}^{a_{6}}(z)[0.7Q_{3}(n)]^{1/2}(q a_{1})^{n+a_{2}}}{1+(q a_{1})^{n+a_{2}}} \nonumber\\
\tilde{b}(n,k) &=& \frac{1+0.2 a_{3} (n+3) (q a_{7})^{n+3+a_{8}}}{1+(q a_{7})^{n+3.5+a_{8}}} \nonumber\\
\tilde{c}(n,k) &=& \frac{1+4.5 a_{4}/[1.5+(n+3)^{4}] (q a_{5})^{n+3+a_{9}}}{1+(q a_{5})^{n+3.5+a_{9}}},
\label{eq:abc}
\end{eqnarray}

\noindent
where $ n $ is the slope of the linear matter power spectrum at wavenumber k and it is given by \citep{LSS10}

\begin{equation}
n \equiv \frac{d \hspace{0.1 cm} log \hspace{0.1 cm} P_{lin}(k)}{d \hspace{0.1 cm} log \hspace{0.1 cm} k},
\label{eq:n}
\end{equation}

\noindent
$ Q_{3} $, a function of $ n $ given by \citep{LSS10}

\begin{equation}
Q_{3}(n) = \frac{4-2^{n}}{1+2^{n+1}},
\label{eq:Q3}
\end{equation}

\noindent
$ q $ is defined as $ q \equiv k/k_{nl} $, where $k_{nl}$ is the scale up to which linear matter power spectrum is accurate enough and it is computed by \citep{LSS10}

\begin{equation}
\frac{k_{nl}^{3} P_{lin}(k_{nl})}{2 \pi^{2}} \equiv 1,
\label{eq:knl}
\end{equation}

\noindent
the $ \sigma_{8} $ parameter at a particular redshift given by $ \sigma_{8}(z) = \frac{D_{+}(z)}{D_{+}(z = 0)} \hspace{0.1 cm} \sigma_{8}(z = 0) $ and finally the constants $ a_{1} $ to $ a_{9} $ are given by \citep{LSS10}

\begin{eqnarray}
a_{1} &=& 0.484 \hspace{1 cm} a_{2} = 3.740 \hspace{1 cm} a_{3} = -0.849 \nonumber\\
a_{4} &=& 0.392 \hspace{1 cm} a_{5} = 1.013 \hspace{1 cm} a_{6} = -0.575 \nonumber\\
a_{7} &=& 0.128 \hspace{1 cm} a_{8} = -0.722 \hspace{0.7 cm} a_{9} = -0.926
\label{eq:a1to9}
\end{eqnarray}

\noindent
So, the tree-level matter bi-spectrum in Eq. \eqref{eq:BStree} gets modified through two ways: one is simply replacing linear matter power spectrum with non-linear matter power spectrum and another is through the modification of $ F_{2} $ denoted by $ F_{2}^{eff} $. For more details of the effective matter bi-spectrum, refer to \citep{LSS10}.

In Fig.~\ref{fig:bsequilateral} we plot deviations in the matter bi-spectrum for the cubic Galileon models compared to the $ \Lambda $CDM model for the equilateral configuration ($ k_{1} = k_{2} = k_{3} = k $). For a particular color i.e. for a particular model the continuous and dashed lines correspond to the tree-level (using Eq. \eqref{eq:BStree}) and effective (using Eq. \eqref{eq:BSeff}) matter bi-spectrum respectively. we can see that the changes in the deviations for effective matter bi-spectrum from the tree-level matter bi-spectrum is not significant (upto few percentage level). The deviation in the tree-level (or in the effective) matter bi-spectrum is the most for the $ \epsilon_{i} = 0 $ model and decreases with increasing $ \epsilon_{i} $. The deviation in the tree-level matter bi-spectrum is roughly twice the deviation in the linear matter power spectrum because $ B_{tree} $ contains $ P_{lin}^{2} $ terms in Eq. \eqref{eq:BStree}. Here also all the deviations are negative because cubic Galileon models are always non-phantom.

In Fig.~\ref{fig:bsFldpSqzd} we have plotted percentage deviations in the matter bi-spectrum from the $ \Lambda $CDM model at $ z = 0 $ similar to the top-left panel of the Fig.~\ref{fig:bsequilateral} but for two other triangular configurations. Left and right panels are for the folded ($ k_{1} = 2 k_{2} = 2 k_{3} = k $) and squeezed ($ k_{1} = k_{2} = 20 k_{3} = k $) configurations respectively. The results are almost similar in the folded and squeezed configurations compared to the equilateral configuration.

\section{Weak lensing statistics}

In the context of the structure formation, matter power spectrum or any higher order statistical quantities like matter bi-spectrum, trispectrum etc. are not directly measurable. In literature among the different direct observational aspects, weak lensing statistics is an important application to probe the underlying matter distribution. Weak lensing is one of the strongest probes to the structure formation history of the Universe. Weak gravitational lensing is an intrinsically statistical measurement of the distortion effect of the images of the background galaxies due to the bending of light rays by matter along the line of sight between distant background galaxies and us. This image distortion effect is quantified by an important quantity called convergence. The convergence, $ \kappa $ can probe the underlying matter distribution ($ \delta_{m} $) and it is thus defined as \citep{wl1,wl2,wl3,wl4}

\begin{equation}
\kappa(\hat{n},\chi) = \int_{0}^{\chi} W(\chi') \delta_{m}(\hat{n},\chi') d \chi',
\end{equation}

\noindent
where the integration is along the line of sight with a weight function denoted by $ W(\chi) $. $ \hat{n} $ is the direction in which we measure the distortion effect and $ \chi $ is the comoving distance. Note that the above equation is valid in the weak lensing limit and in the Newtonian perturbation theory. The weight function $ W $ at a particular comoving distance (or at a particular redshift) is given by \citep{wl1,wl2,wl3,wl4}

\begin{equation}
W(\chi(z)) = \frac{3}{2} \Omega_{m}^{(0)} H_{0}^{2} g(z) (1+z),
\end{equation}

\noindent
where $ g $ is given by \citep{wl1,wl2,wl3,wl4}

\begin{equation}
g(z) = \chi(z) \int_{z}^{\infty} dz' n(z') \Big{(} 1-\frac{\chi(z)}{\chi(z')} \Big{)},
\end{equation}

\noindent
In literature $ g(z)/\chi(z) $ is called the geometric lensing efficiency factor. Here $ n(z) $ is the source distribution function. It is normalized by the condition $ \int_{0}^{\infty} n(z) dz = 1 $. We consider a particular source distribution given by \citep{LSSextra2,wl1,wl2,wl3,wl4}

\begin{equation}
n(z) = \frac{(\frac{1 + b_{1}}{z_{0}})}{\Gamma \Big{(} \frac{1+b_{1} + b_{2}}{b_{2}} \Big{)} } \Big{(} \frac{z}{z_{0}} \Big{)}^{b_{1}} \exp{\Big{[} - \Big{(} \frac{z}{z_{0}} \Big{)}^{b_{2}}} \Big{]},
\label{eq:nOfz}
\end{equation}

\noindent
where the normalization condition $ \int_{0}^{\infty} n(z) dz = 1 $ has been used. And we consider $ b_{1} = 2 $, $ b_{2} = 1.5 $ and $ z_{0} = 0.9/1.412 $ which are similar to the Euclid Survey \citep{euclid2,euclid3,nz,pwl1}.

\subsection{Convergence power spectrum}

The convergence power spectrum, $ P_{\kappa} $ is defined as (assuming statistical isotropy)

\begin{equation}
< \kappa_{l m} \kappa_{l' m'}^{*} > = \delta_{l l'} \delta_{m m'} P_{\kappa}(l),
\end{equation}

\noindent
where $ \kappa_{lm} $ is the fourier transform of $ \kappa(\hat{n},\chi) $ in the multipole ($ l $,$ m $) space and it is given by

\begin{equation}
\kappa_{lm} = \int d \hat{n} \kappa(\hat{n},\chi) Y_{lm}^{*},
\end{equation}

\noindent
where $ Y_{lm} $ are the spherical harmonics. In the Newtonian perturbation theory and using the Limber approximation the convergence power spectrum becomes \citep{wl1,wl2,wl3,wl4,wl5,wl6}

\begin{equation}
P_{\kappa}(l) = \int_{0}^{\infty} \frac{d z}{H(z)} \frac{W^{2}(\chi(z))}{\chi^{2}(z)} P \Big{(} \frac{l}{\chi(z)},z \Big{)}.
\label{eq:PSkappa}
\end{equation}

%%%%%%%%%%%%%%%%%%%%%%%%%%%%%%%%%%
\begin{figure}[tbp]
\centering
\includegraphics[width=.7\textwidth]{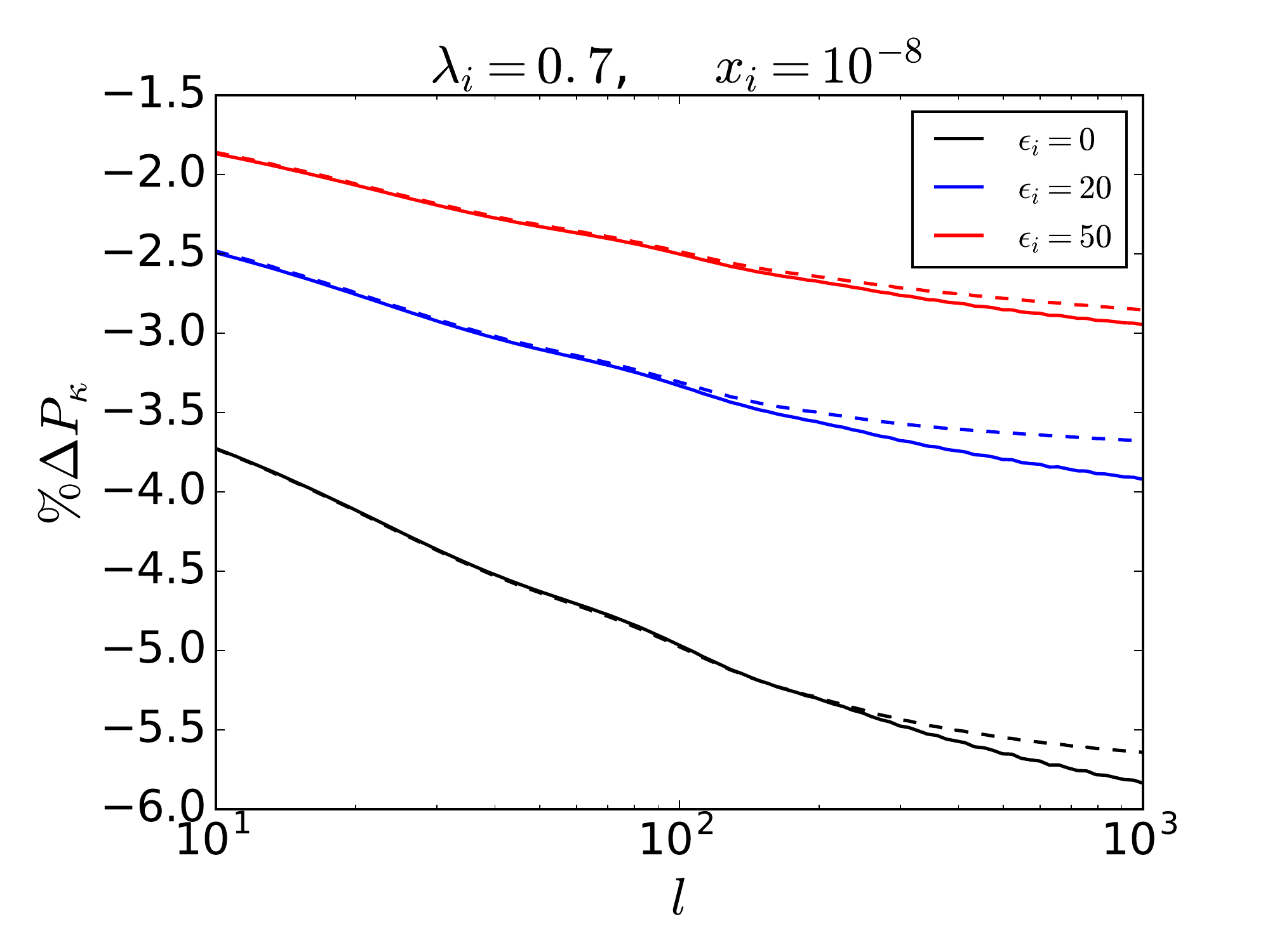}
\caption{\label{fig:wlps} Percentage deviation in convergence power spectrum from the $ \Lambda $CDM model. Continuous lines are for the results putting linear matter power spectrum of Eq. \eqref{eq:PSlinear} into Eq. \eqref{eq:PSkappa}. Dashed lines are for the results putting the non-linear matter power spectrum from HMCode \citep{HMCode} into Eq. \eqref{eq:PSkappa}. All the models have the negative deviations as because all the models are non-phantom. The deviation (negative deviation) is the largest for the quintessence model and the deviations (negative deviations) increase with increasing $\epsilon_{i}$.}
\end{figure}
%%%%%%%%%%%%%%%%%%%%%%%%%%%%%%%%%%

%%%%%%%%%%%%%%%%%%%%%%%%%%%%%%%%%%
\begin{figure}[tbp]
\centering
\includegraphics[width=.495\textwidth]{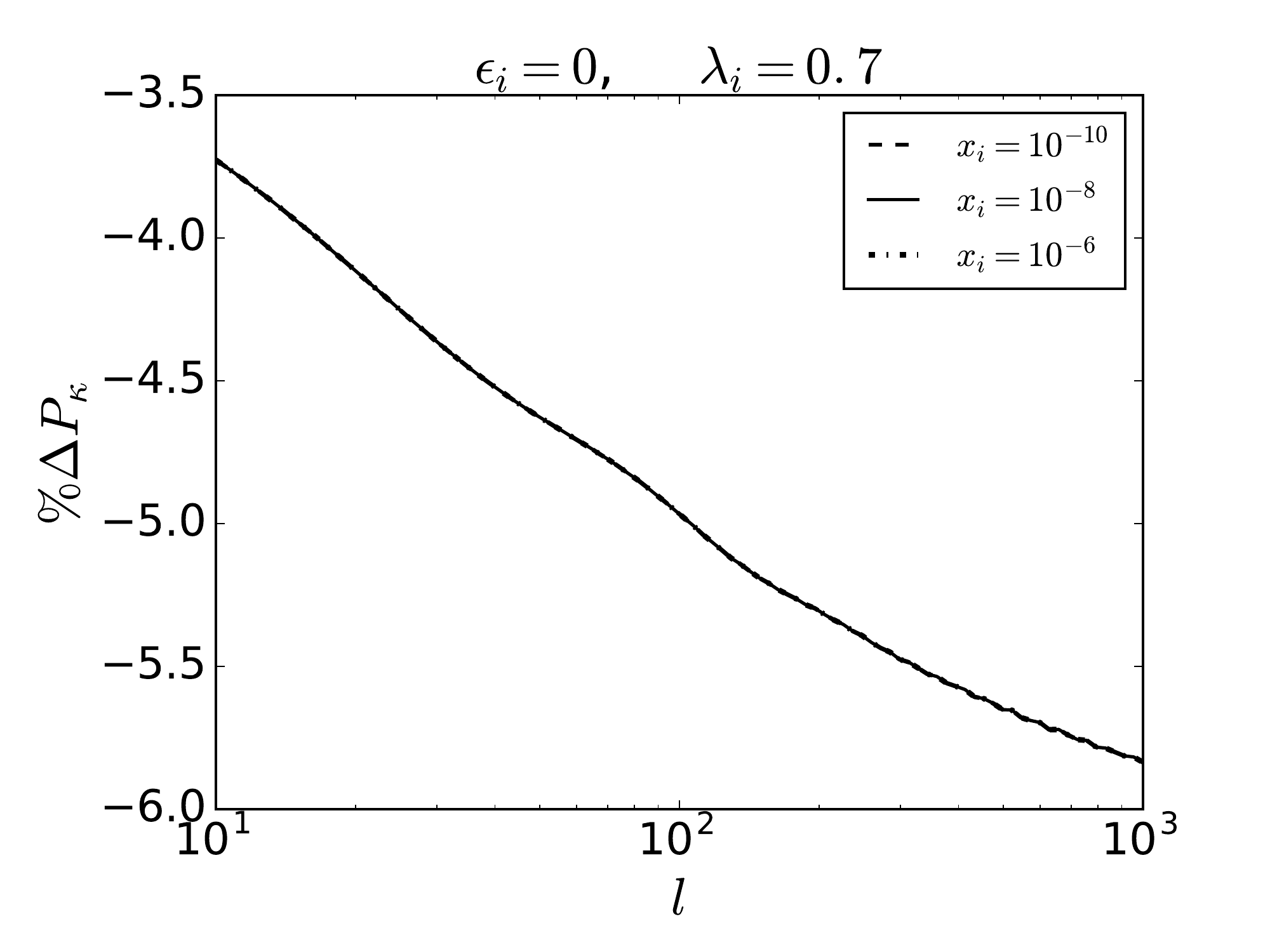}
\includegraphics[width=.495\textwidth]{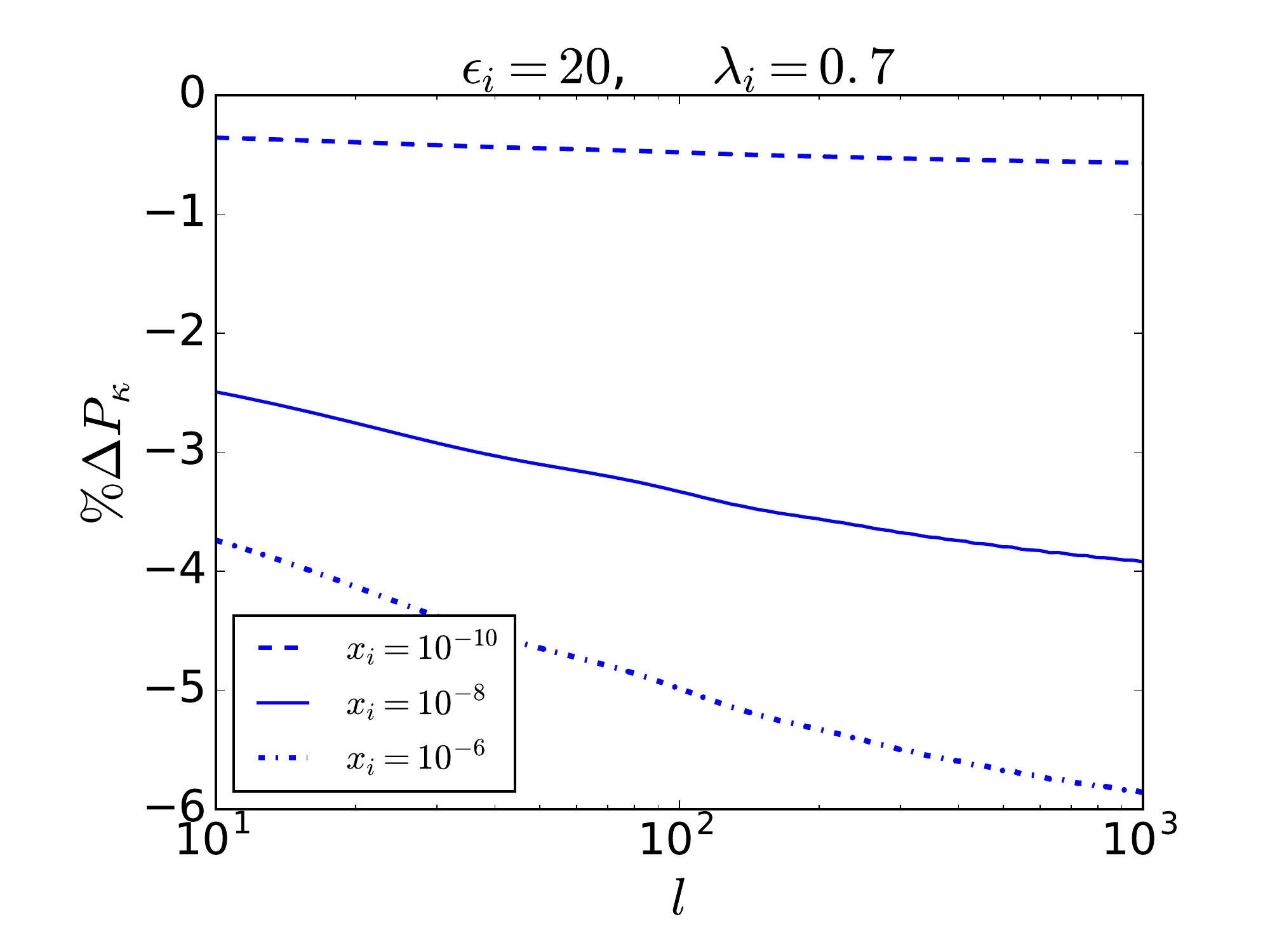}
\includegraphics[width=.495\textwidth]{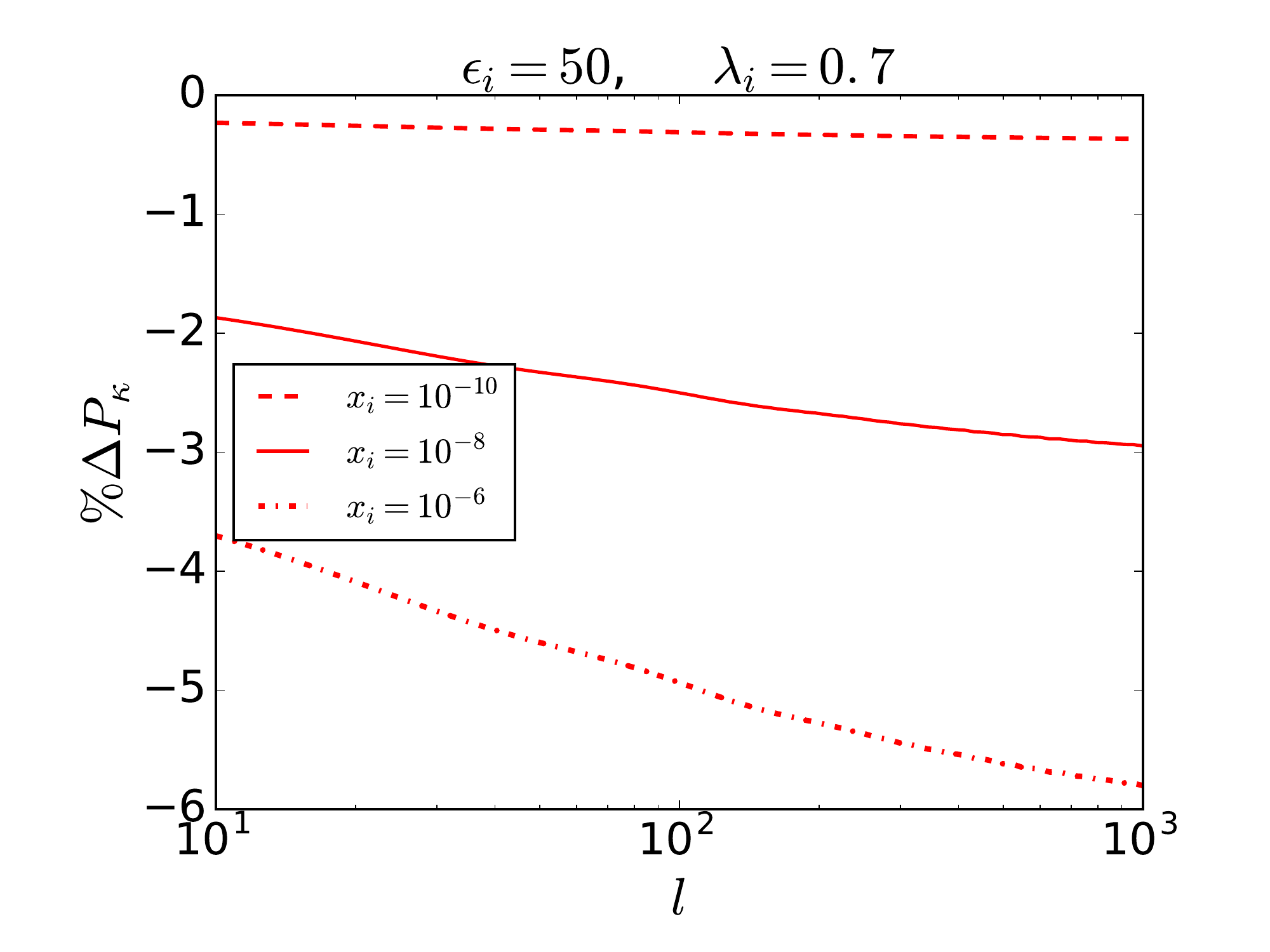}
\caption{\label{fig:wlps2} Percentage deviation in convergence power spectrum from the $ \Lambda $CDM model with the variation of the $x_{i}$ parameter. Continuous lines are for the results putting linear matter power spectrum of Eq. \eqref{eq:PSlinear} into Eq. \eqref{eq:PSkappa}. Dashed lines are for the results putting the non-linear matter power spectrum from HMCode \citep{HMCode} into Eq. \eqref{eq:PSkappa}. The dependency of the deviations on the $x_{i}$ parameter is the smallest for the quintessence model (in fact negligible) and increases with increasing $\epsilon_{i}$. The deviations (negative deviations) increase with increasing $x_{i}$ from the $\Lambda$CDM model.}
\end{figure}
%%%%%%%%%%%%%%%%%%%%%%%%%%%%%%%%%%

%%%%%%%%%%%%%%%%%%%%%%%%%%%%%%%%%%
\begin{figure}[tbp]
\centering
\includegraphics[width=.495\textwidth]{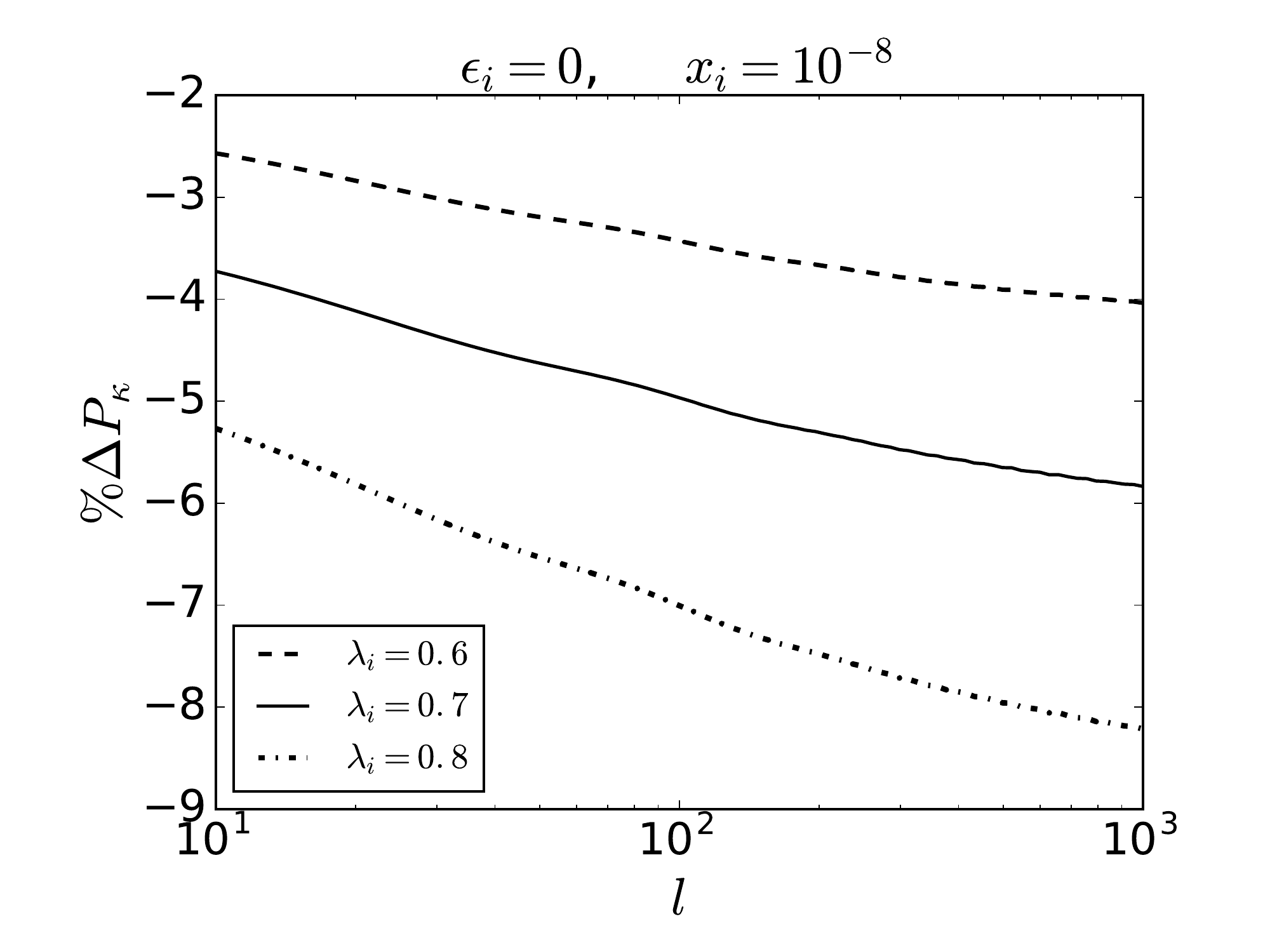}
\includegraphics[width=.495\textwidth]{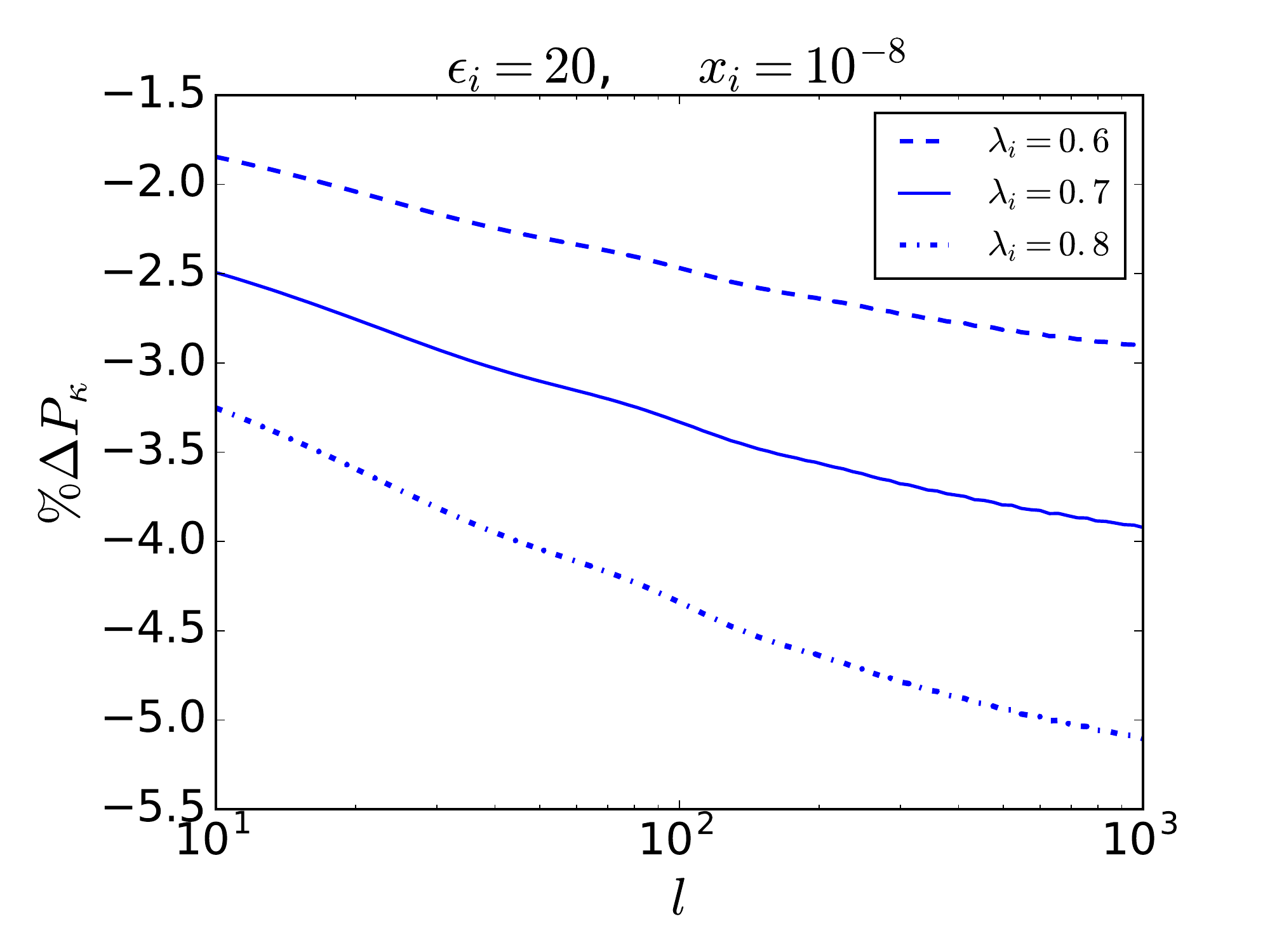}
\includegraphics[width=.495\textwidth]{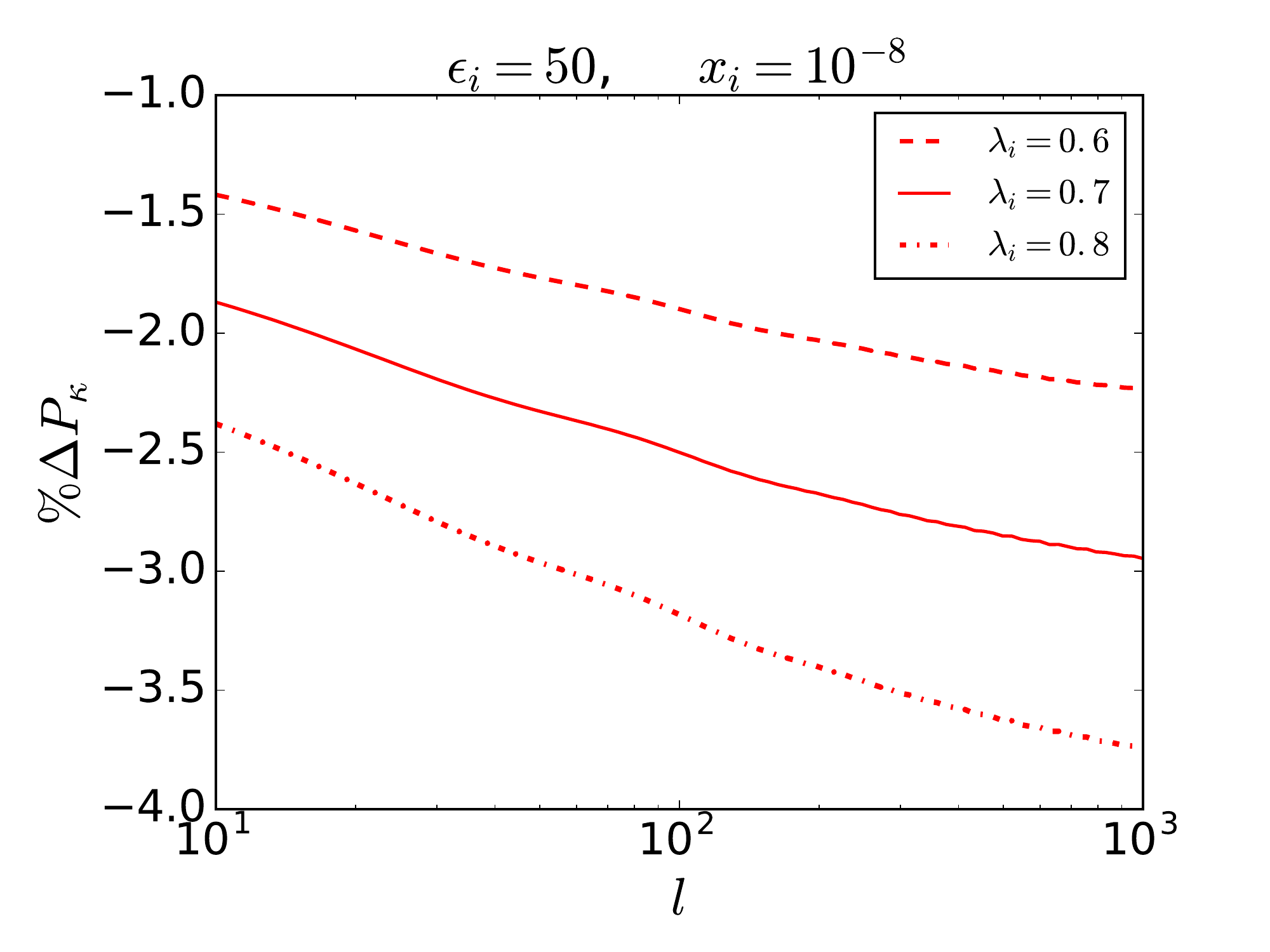}
\caption{\label{fig:wlps3} Percentage deviation in convergence power spectrum from the $ \Lambda $CDM model with the variation of the $\lambda_{i}$ parameter. Continuous lines are for the results putting linear matter power spectrum of Eq. \eqref{eq:PSlinear} into Eq. \eqref{eq:PSkappa}. Dashed lines are for the results putting the non-linear matter power spectrum from HMCode \citep{HMCode} into Eq. \eqref{eq:PSkappa}. The dependency of the deviations on the $\lambda_{i}$ parameter is the largest for the quintessence model and decreases with increasing $\epsilon_{i}$. The deviations (negative deviations) increase with increasing $\lambda_{i}$ from the $\Lambda$CDM model.}
\end{figure}
%%%%%%%%%%%%%%%%%%%%%%%%%%%%%%%%%%

In Fig.~\ref{fig:wlps} we have plotted percentage deviations in the convergence power spectrum for the cubic Galileon models from the $ \Lambda$CDM model. The same color corresponds to the same value of $ \epsilon_{i} $ whereas continuous and dashed lines correspond to the computation of the convergence power spectrum by using linear and non-linear power spectrum respectively in Eq. \eqref{eq:PSkappa}. As because cubic Galileon models are always non-phantom, the deviations here are always negative similar to the previous results. The deviation is the largest for the $ \epsilon_{i} = 0 $ model. The deviation decreases with increasing $ \epsilon_{i} $. Note that the deviations increase with increasing $ l $. This is because of the fact that different dark energy models have different comoving distance at a fixed redshift. So, in the argument of the power spectrum same k corresponds to the different $ l $ at a particular redshift for different dark energy models. For this reason, the scale dependency arises in the deviations in the convergence power spectrum although there is no scale dependency in the deviations in the matter power spectrum. Now, smaller comoving distance corresponds to lesser the intervening matter components and hence lesser the intervening matter energy density. Hence weak lensing signal reduces. Since cubic Galileon model is always non-phantom the weak lensing signal is less in this model compared to the $ \Lambda$CDM model. So, the deviations in the convergence power spectrum increase in the negative y-axis with increasing $ l $. We can notice that the deviations between linear and non-linear results are accurate enough (less than a percent) up to a scale $l = 10^{3}$.

Note that the deviations in the background (for example Hubble parameter) as well as the perturbation (for example matter power spectrum) quantities in each value of redshift added up in the deviations of the convergence power spectrum. This can be seen from the fact that the integration in Eq. \eqref{eq:PSkappa} with respect to the redshift can be considered as the summation of the integrand over all the redshift bins with very small redshift interval. So, the deviations in the convergence power spectrum significantly increase from the deviations in the Hubble parameter, matter power spectrum (etc at a fixed redshift). This argument is also true for the convergence bispectrum which will be discussed in the next subsection.

In Figs.~\ref{fig:wlps2} and ~\ref{fig:wlps3}, we have shown the $x_{i}$ and $\lambda_{i}$ parameters dependency on the convergence power spectrum. From Figs.~\ref{fig:Hubble} and ~\ref{fig:wlps}, we know that the positive deviations in the Hubble parameters correspond to the negative deviation in the convergence power spectrum. More the positive deviations in the Hubble parameters more the negative deviation in the convergence power spectrum. This fact is reflected in Figs.~\ref{fig:wlps2} and ~\ref{fig:wlps3} according to the Figs.~\ref{fig:Hubble2} and ~\ref{fig:Hubble3} respectively.

To summarize, the convergence power spectrum decreases with increasing $x_{i}$ (except $\epsilon_{i} = 0$). So, the deviations increase with increasing $x_{i}$ from the $\lambda$CDM model. For the quintessence model, convegence power spectrum hardly depend on the $x_{i}$ parameters.

Similarly, the convergence power spectrum decreases with increasing $\Lambda_{i}$ for all the cubic Galileon models. This corresponds to the fact that the deviations in the convergence power spectrum increase with increasing values of $\lambda_{i}$ from the $\Lambda$CDM model. However, one interesting point is that the dependency of the deviations on the $\lambda_{i}$ parameter decreases with the increasing values of $\epsilon_{i}$.

\subsection{Convergence bi-spectrum}

The convergence bi-spectrum is defined as \citep{wl1,wl3,wl7,wl8}

\begin{equation}
< \kappa_{l_{1} m_{1}} \kappa_{l_{2} m_{2}} \kappa_{l_{3} m_{3}} > = \left( \begin{smallmatrix} l_{1} \hspace{0.5 cm} & l_{2} & \hspace{.5 cm} l_{3} \vspace{0.5 cm} \\ m_{1} \hspace{.5 cm} & m_{2} & \hspace{.5 cm} m_{3} \end{smallmatrix} \right) B_{l_{1} l_{2} l_{3}}^{\kappa},
\end{equation}

\noindent
where $ \left( \begin{smallmatrix} l_{1} \hspace{0.2 cm} & l_{2} & \hspace{0.2 cm} l_{3} \vspace{0.2 cm} \\ m_{1} \hspace{0.2 cm} & m_{2} & \hspace{0.2 cm} m_{3} \end{smallmatrix} \right) $ is the Wigner 3-j symbol. In the above definition $ B_{l_{1} l_{2} l_{3}}^{\kappa} $ is the full sky convergence bi-spectrum and it is related to the flat sky convergence bi-spectrum given by \citep{wl7}

\begin{equation}
B_{l_{1} l_{2} l_{3}}^{\kappa} = \left( \begin{smallmatrix} l_{1} \hspace{0.5 cm} & l_{2} & \hspace{.5 cm} l_{3} \vspace{0.5 cm} \\ 0 \hspace{.5 cm} & 0 & \hspace{.5 cm} 0 \end{smallmatrix} \right) \sqrt{\frac{(2 l_{1} + 1) (2 l_{2} + 1) (2 l_{3} + 1)}{4 \pi}} B_{\kappa}(\vec{l}_{1},\vec{l}_{2},\vec{l}_{3};z).
\end{equation}

\noindent
In the Newtonian perturbation theory and using the Limber approximation, convergence bi-spectrum becomes \citep{wl1,wl3,wl7,wl8}

\begin{equation}
B_{\kappa}(\vec{l}_{1},\vec{l}_{2},\vec{l}_{3};z) = \int_{0}^{\infty} \frac{d z}{H(z)} \frac{W^{3}(\chi(z))}{\chi^{4}(z)} B \Big{(} \frac{\vec{l}_{1}}{\chi(z)},\frac{\vec{l}_{2}}{\chi(z)},\frac{\vec{l}_{3}}{\chi(z)};z \Big{)},
\label{eq:BSkappaflatsky}
\end{equation}

\noindent
with $ \vec{l}_{1} + \vec{l}_{2} + \vec{l}_{3} = 0 $.

%%%%%%%%%%%%%%%%%%%%%%%%%%%%%%%%%%
\begin{figure}[tbp]
\centering
\includegraphics[width=.495\textwidth]{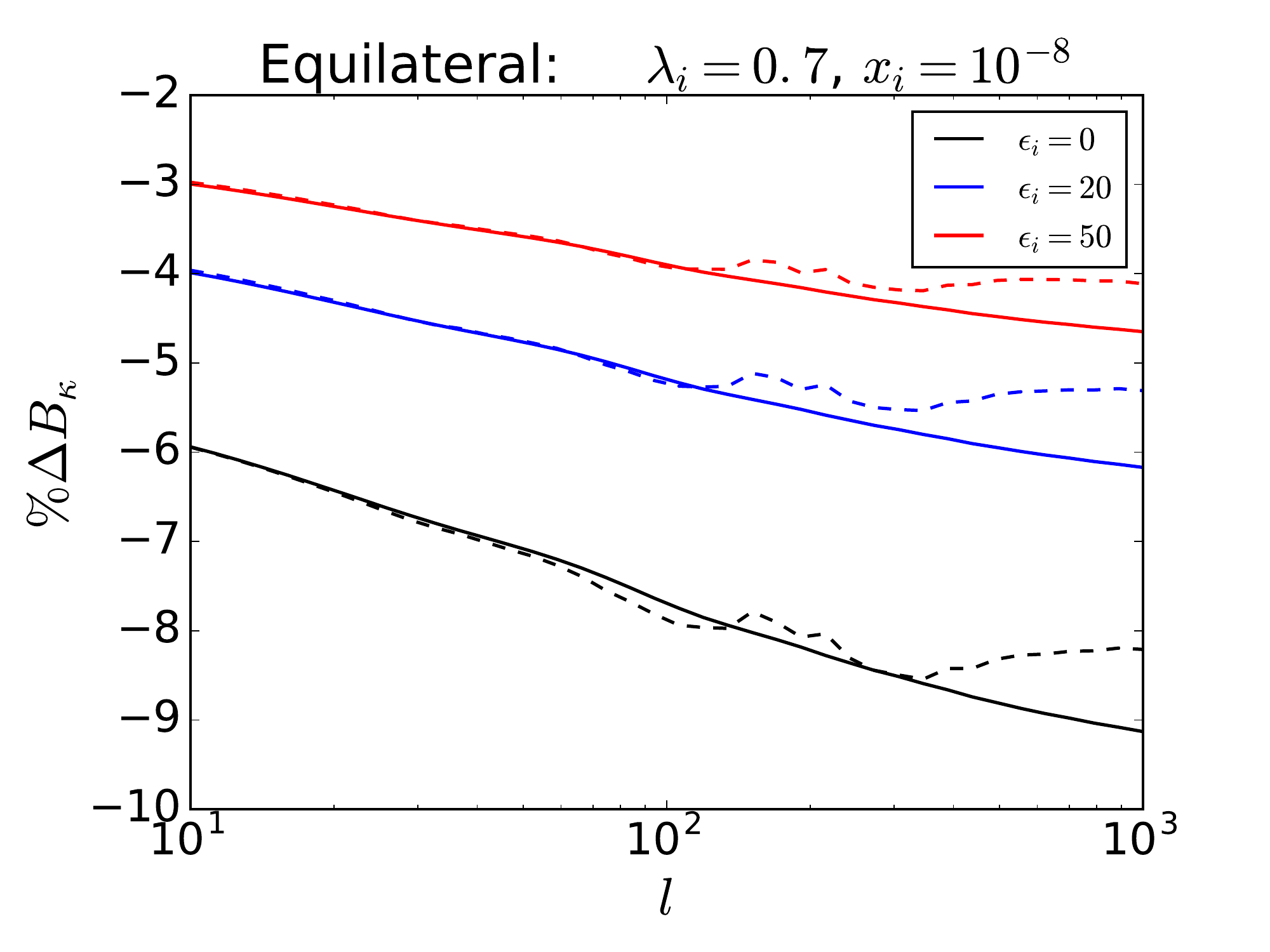}
\includegraphics[width=.495\textwidth]{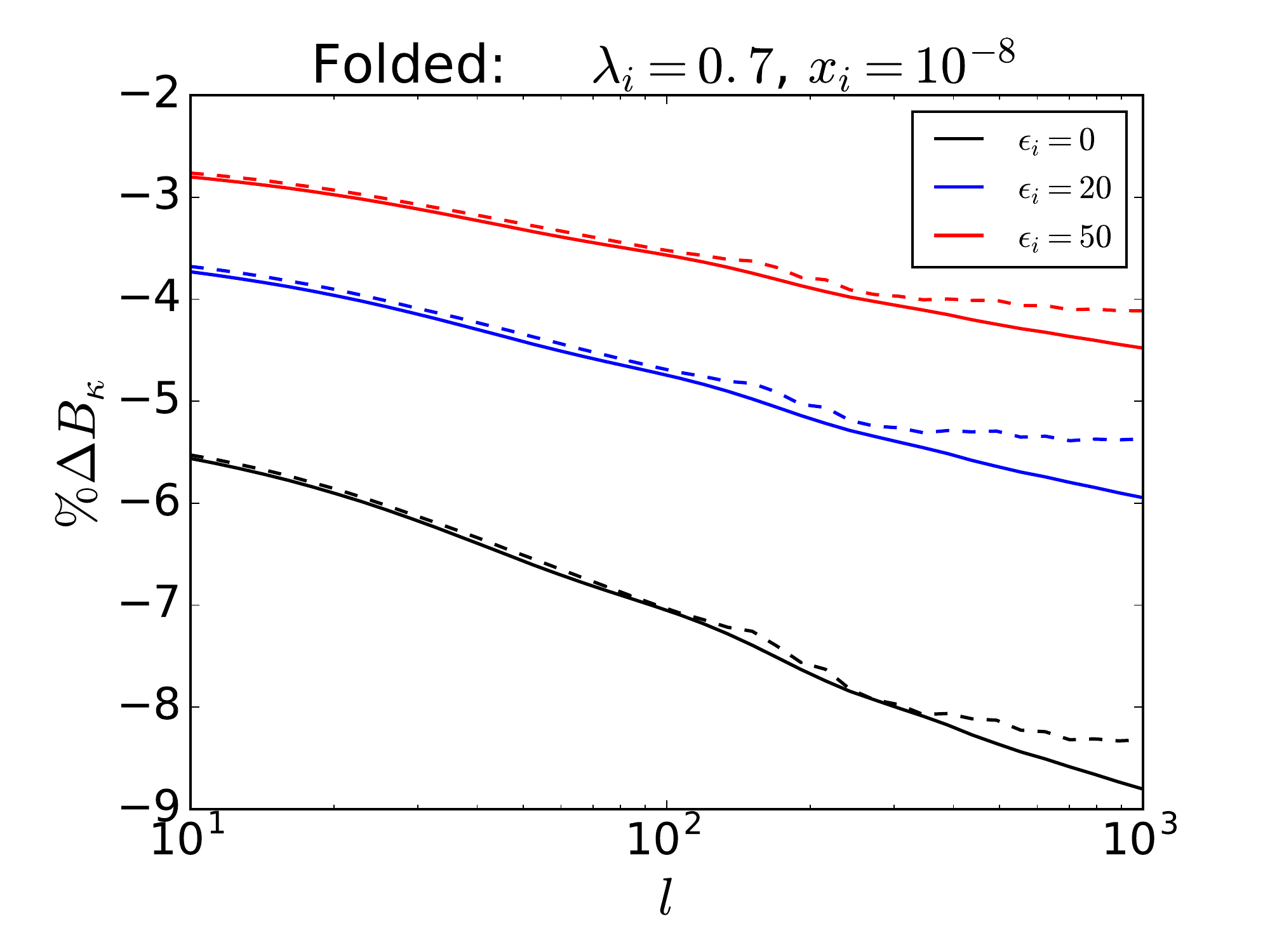}\\
\includegraphics[width=.495\textwidth]{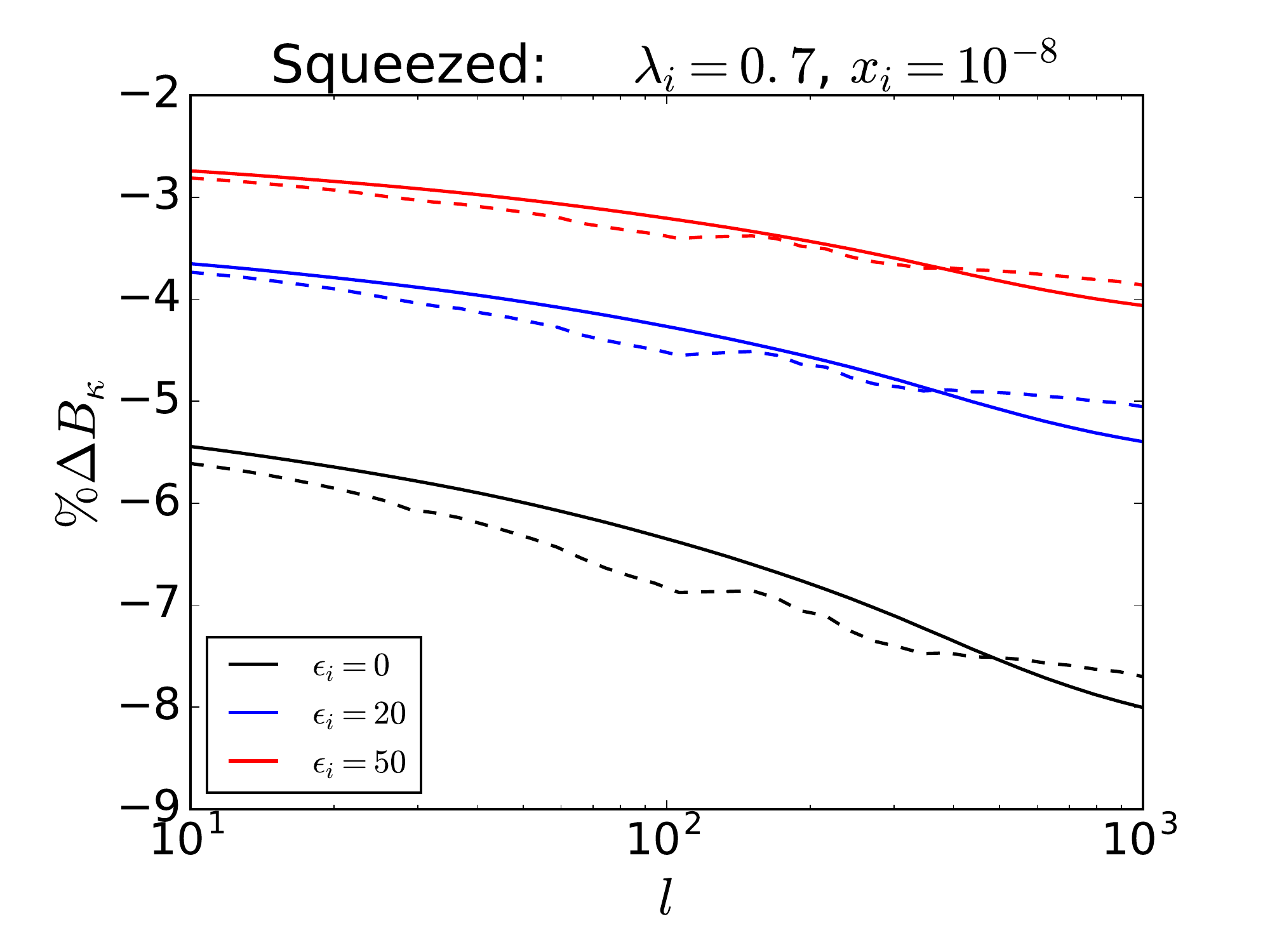}
\caption{\label{fig:wlbs} Percentage deviation in convergence bi-spectrum from the $ \Lambda $CDM model. Continuous lines are for the results putting tree-level matter bi-spectrum of Eq. \eqref{eq:BStree} into Eq. \eqref{eq:BSkappaflatsky}. Dashed lines are for the results putting the effective non-linear matter bi-spectrum of Eq. \eqref{eq:BSeff} into Eq. \eqref{eq:BSkappaflatsky}. The behavior of the deviations is similar as in Fig.~\ref{fig:wlps}, only the amplitude of the deviations increases. All the three configurations have almost similar deviations.}
\end{figure}
%%%%%%%%%%%%%%%%%%%%%%%%%%%%%%%%%%

In Fig.~\ref{fig:wlbs} percentage deviations in the convergence bi-spectrum from the $ \Lambda $CDM model has been plotted using equation \eqref{eq:BSkappaflatsky} for the equilateral ($ l_{1} = l_{2} = l_{3} = l $), folded ($ l_{1} = 2 l_{2} = 2 l_{3} = l $) and squeezed ($ l_{1} = l_{2} = 20 l_{3} = l $) configurations respectively. First of all note that since we have compared the results between two dark energy models, the results are exactly same for both the flat sky and full sky convergence bi-spectrum. The same color corresponds to the same value of $ \epsilon_{i} $. The continuous lines correspond to the results of using the tree-level matter bi-spectrum of Eq. \eqref{eq:BStree} in Eq. \eqref{eq:BSkappaflatsky}. The dashed lines correspond to the results of using the effective non-linear matter bi-spectrum of Eq. \eqref{eq:BSeff} in Eq. \eqref{eq:BSkappaflatsky}. The results are almost similar in the three configurations. Since the cubic Galileon models are always non-phantom, the deviations are always negative as consistant to the previous results. The deviations in the convergence bi-spectrum increase with increasing $ l $ because of the same reason mentioned before (in the discussion of the deviations in the convergence power spectrum). The deviation is the most for $ \epsilon_{i} = 0 $ and decreases with increasing $ \epsilon_{i} $. Here also we can see that the linear and non-linear results are less than percentage level up to $ l = 10^{3} $ scale.

We have not explicitly shown the dependency of the convergence bispectrum on $x_{i}$ and $\lambda_{i}$ parameters as because of the dependency will be similar as in the case of the convergence power spectrum. Only the amplitudes of the deviations in the convergence bispectrum increases compared to the deviations in the convergence power spectrum (which can be seen from the above discussions).

\subsection{Chi-square analysis}

The statistical error in the convergence power spectrum is given by \citep{WLp6,WLp11a,WLp11}

\begin{equation}
\Delta P_{\kappa} (l) = \sqrt{\frac{2}{(2l+1)f_{sky}}} \left[ P_{\kappa} (l) +\frac{<\gamma_{int}^{2}>}{\bar{n}} \right],
\label{eq:errorPwl}
\end{equation}

\noindent
where $f_{sky}$ is a fraction of the sky covered by a given survey. In the right-hand side, the last term corresponds to the shot noise term. $<\gamma_{int}^{2}>^{1/2}$ corresponds to the intrinsic galaxy ellipticity. $\bar{n}$ is the average number of galaxies per steradian for the ellipticity measurement. $\bar{n}$ can be expressed as $\bar{n}=3600(\frac{180}{\pi})^{2} n_{\theta}$ \citep{pwlnew1,pwlnew2,pwlnew3}, where $n_{\theta}$ being the total number of galaxies per $arcmin^{2}$. Here, we consider the Euclid configurations given by $f_{sky}=0.364$ \citep{pwlnew1,euclid2}, $<\gamma_{int}^{2}>^{1/2}=0.22$ \citep{pwlnew1,pwlnew2,pwlnew3}, and $n_{\theta}=30$ citep{pwlnew1,pwlnew2}.

%\noindent
From Eq. \eqref{eq:errorPwl}, we can calculate the theoretical error associated with the convergence power spectrum for a particular model in a particular survey. Now, we take $\Lambda$CDM as a fiducial model and compute the error associated with the convergence power spectrum in $l$-bins. From Fig.~\ref{fig:wlps}, we have seen that the deviations in the convergence power spectrum between different models are less than $1\%$ for nonlinear vs. linear results. So, to make $l$-bins, we restrict our analysis to $l_{max}=1000$ to work in the linear theory during the $\chi^{2}$-analysis. For the minimum value, we put $l_{min}=10$. We avoid the regime $l<10$ to reduce the effect of the cosmic variance in the $\chi^{2}$-analysis. For the $l$-bins, we consider $\Delta \log l = 0.2$ \citep{euclid2}. These $l$ intervals correspond to the $24$ $l$-bins in the regime $10 \leq l \leq 1000$. Finally, we compute simulated data for the errors in the convergence power spectrum for each $l$-bin using $\Lambda$CDM as the fiducial model. Now, we define $\chi^{2}$ as

\begin{equation}
\chi^{2} (\epsilon_{i},\lambda_{i},x_{i}) = \sum_{l=l_{min}}^{l_{max}} \frac{\left[P_{\kappa}^{gal}(l,\epsilon_{i},\lambda_{i},x_{i})-P_{\kappa}^{\Lambda CDM}(l) \right]^{2}}{\left[\Delta P_{\kappa}^{\Lambda CDM}(l) \right]^{2}},
\label{eq:chiSquare}
\end{equation}

\noindent
where $\epsilon_{i}$, $\lambda_{i}$, and $x_{i}$ are the cubic Galileon model parameters, discussed in the text. The superscripts 'gal' and '$\Lambda$CDM' correspond to the cubic Galileon and $\Lambda$CDM models respectively. We use publicly available codes PyMultinest (Multinest algorithm \citep{multinest1,multinest2}) and getdist plot to draw the $1$-$\sigma$ and $2$-$\sigma$ contour plots from Eq. \eqref{eq:errorPwl}. To do the $\chi^2{}$-analysis, we vary the parameters in the ranges given by $0 \leq \epsilon_{i} \leq 100$, $0 \leq \lambda_{i} \leq 1$, and $10^{-10} \leq x_{i} \leq 10^{-2}$ respectively.

%%%%%%%%%%%%%%%%%%%%%%%%%%%%%%%%%%
\begin{figure}[tbp]
\centering
\includegraphics[width=0.9\textwidth]{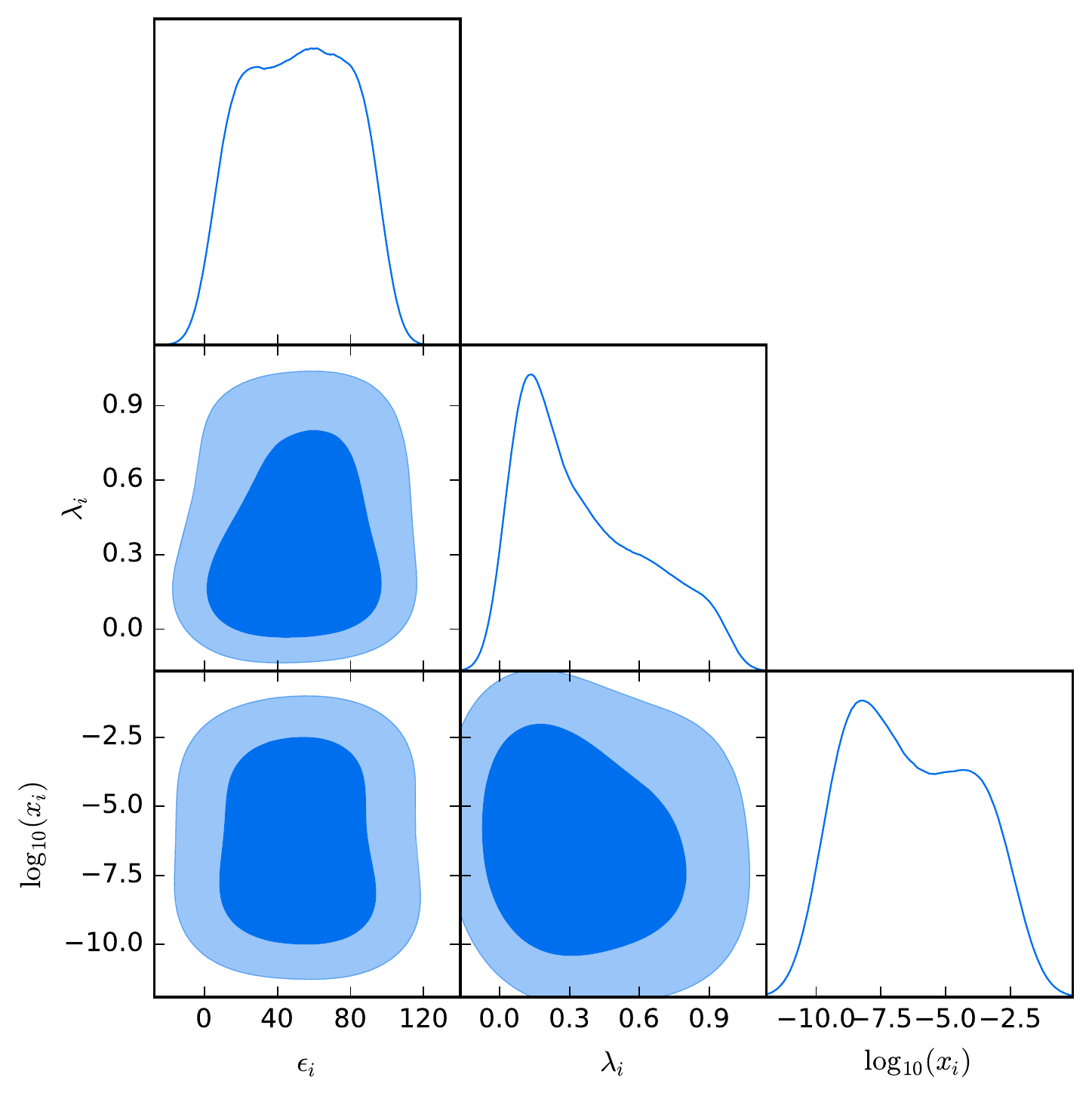}
\caption{\label{fig:trianglePlot} Triangle plot. $1$-$\sigma$ and $2$-$\sigma$ confidence contours in the triangle plot in all the three parameters ($\epsilon_{i}$, $\lambda_{i}$, and $x_{i}$ (actually $\log_{10}(x_{i})$)) from the $\chi^{2}$-analysis given by Eq. \eqref{eq:chiSquare}. The main result is that the higher values of $\lambda_{i}$ are less allowed as we take $\Lambda$CDM as the fiducial model.}
\end{figure}
%%%%%%%%%%%%%%%%%%%%%%%%%%%%%%%%%%

%%%%%%%%%%%%%%%%%%%%%%%%%%%%%%%%%%
\begin{figure}[tbp]
\centering
\includegraphics[width=.49\textwidth]{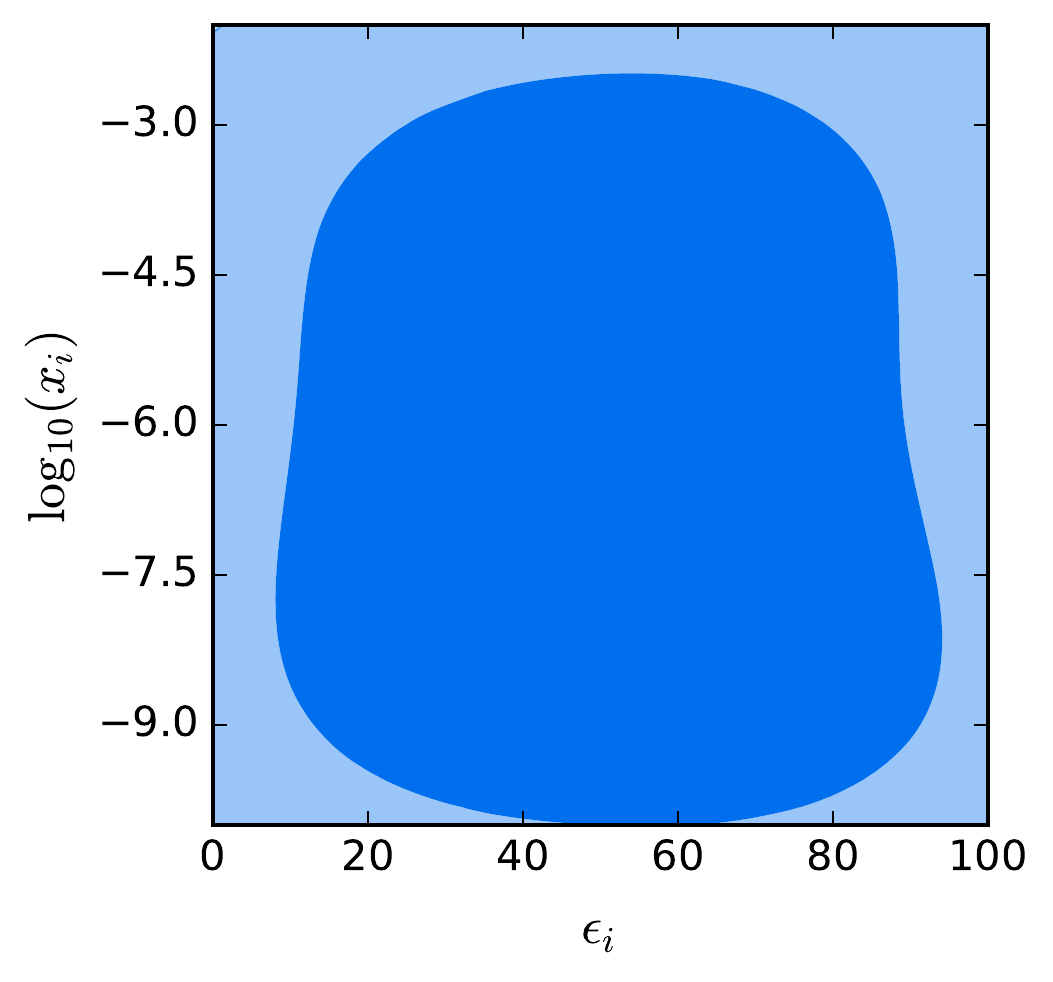}
\includegraphics[width=.49\textwidth]{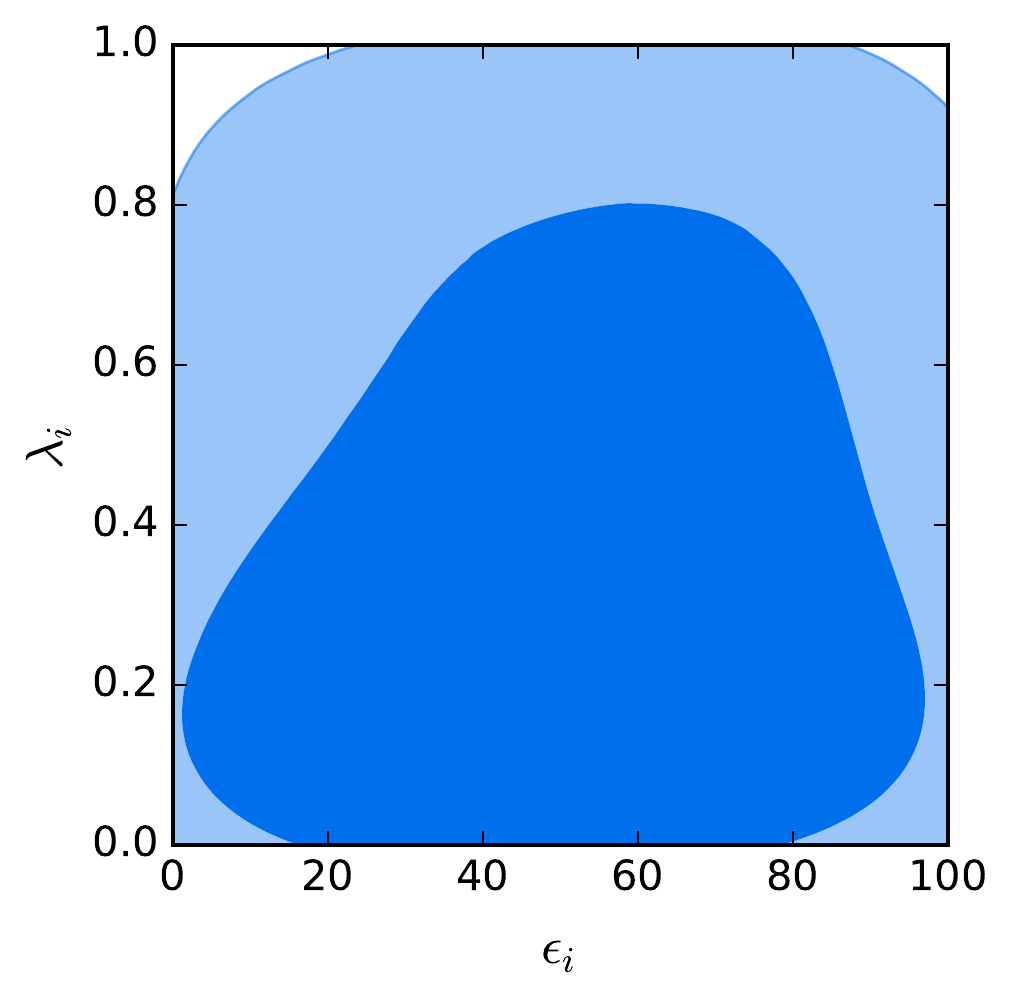}
\caption{\label{fig:contourPlots} Contour plots from the triangle plot. Same as in Fig.~\ref{fig:trianglePlot}. We have separated out the two plots (among the three plots) to show the behaviors explicitly. From the left panel, we can see that for $x_{i}<<1$, all the $x_{i}$ values are in the $1$-$\sigma$ contour and only the values $\epsilon_{i}<5$ are outside the $1$-$\sigma$ contour. Only the very large values of $x_{i}$ (comparable to $1$) are outside the $1$-$\sigma$ contour. The right panel has an interesting behavior: the region with larger $\lambda_{i}$ and smaller $\epsilon_{i}$ are significantly outside the $1$-$\sigma$ contour. So, in this region, quintessence and cubic Galileon model can be significantly distinguished.}
\end{figure}
%%%%%%%%%%%%%%%%%%%%%%%%%%%%%%%%%%

\noindent
In Fig.~\ref{fig:trianglePlot}, we have shown the triangle plot with the $1$-$\sigma$ and $2$-$\sigma$ confidence intervals for the three parameters $\epsilon_{i}$, $\lambda_{i}$, and $x_{i}$ respectively. In Fig.~\ref{fig:contourPlots}, we have plotted the same confidence intervals as in Fig.~\ref{fig:trianglePlot} to show the contours seperately with zoomed in version. 

%\noindent
From the left panel of  Fig.~\ref{fig:contourPlots} it is clear that the confidence regions do not depend on $x_{i}$ significantly as long as $x_{i} \ll 1$. However, the higher values of $x_{i}$ (where $x_{i}$ is comparable to $1$) are out of the $1$-$\sigma$ confidence contour. This is because as we increase the value of $x_{i}$, the cubic Galileon model deviates more from the $\Lambda$CDM model (except $\epsilon=0$). This result is consistent with Figs.~\ref{fig:Hubble2} and ~\ref{fig:wlps2}. One other point to notice that $\epsilon<5$ region is outside of the $1$-$\sigma$ contour.

%\noindent
The right panel of  Fig.~\ref{fig:contourPlots} is comparatively more important. In this plot, we can see that the models with higher values of $\lambda_{i}$ are outside of the $1$-$\sigma$ contour for the lower values of $\epsilon_{i}$. So, the quintessence model can be significantly distinguished from the cubic Galileon model for the higher values of $\lambda_{i}$ (at least by the $1$-$\sigma$ confidence). This is because as we increase the value of $\lambda_{i}$, the quintessence model deviates more from the $\Lambda$CDM model compared to the cubic Galileon model. This result is consistent with Figs.~\ref{fig:Hubble3} and ~\ref{fig:wlps3}. So, larger the value of $\lambda_{i}$ smaller the allowed region for the models with smaller $\epsilon_{i}$ values.

%\noindent
So, in conclusion, the quintessence, and the cubic Galileon models can be distinguished in $1$-$\sigma$ confidence by only a simple $\chi^{2}$-analysis. However, the full $\chi^{2}$-analysis with tomographic weak lensing statistics and with the other data can put stringent constraints on the $\epsilon_{i}$ parameter.

\section{Conclusion}

We have considered lowest non-trivial order action of the full Galileon action i.e. the cubic Galileon model to study the late-time cosmic acceleration. We have studied the signature of the cubic Galileon model on the growth of the structures through the convergence power spectrum and bi-spectrum.

The deviations in the convergence power spectrum are 2.5 to 5$ \% $ at $ l = 10^{2} $ for the cubic Galileon models from the $ \Lambda $CDM model. The deviations increase with increasing $ l $. To compute the convergence power spectrum we have considered both the linear and non-linear matter power spectrum and the differences in the results are less than 1$ \% $ up to $ l = 10^{3} $. The linear results are very accurate up to the scale $ l \approx 10^{2} $.

The deviations in the convergence bi-spectrum are 4 to 8$ \% $ at $ l = 10^{2} $ for the cubic Galileon models from the $ \Lambda $CDM model for the equilateral configuration. The results are almost similar for the folded and squeezed configurations. The deviations increase with increasing $ l $. To compute the convergence bi-spectrum we have considered both the tree-level and the effective non-linear matter bi-spectrum and the differences in the two results are sub-percentage up to the scale $ l = 10^{3} $.

The quintessence model can be significantly distinguished from the cubic Galileon model for the higher values of $\lambda_{i}$ (higher values of the initial slopes of the potential).

In summary, future high precision cosmological observations like DES, LSST, Euclid, WFIRST etc. can decisively detect the cubic Galileon models from the $ \Lambda $CDM model or from the quintessence models through the measurements of the convergence power spectrum and bi-spectrum.

\acknowledgments

The author would like to acknowledge Council of Scientific $\&$ Industrial Research (CSIR), Govt. of India for financial support through Senior Research Fellowship (SRF) scheme No:09/466(0157)/2012-EMR-I.

%\paragraph{Note added.} This is also a good position for notes added
%after the paper has been written.


\end{document}